\documentclass[aps,prl,longbibliography,twocolumn,superscriptaddress]{revtex4-2}
\usepackage{amsmath,amssymb,bm,graphicx,color,gensymb,bbold,hyperref,keyval,url,latexsym}
\usepackage{CJK}
\usepackage[dvipsnames,svgnames,table]{xcolor}
\usepackage{enumitem}

\usepackage{hyperref}
\hypersetup{
    pdfstartview={FitH},
    colorlinks=true,    
    linkcolor=NavyBlue, 
    citecolor=Maroon,   
    filecolor=NavyBlue, 
    urlcolor=NavyBlue   
}

\begin{document}

\title{Phase Diagram of the Easy-Axis Triangular-Lattice $J_1$--$J_2$ Model}

\author{Cesar A. Gallegos}
\affiliation{Department of Physics and Astronomy, University of California, Irvine, California 92697, USA}
\begin{CJK*}{UTF8}{}
\author{Shengtao Jiang (\CJKfamily{gbsn}蒋晟韬)}
\affiliation{Department of Physics and Astronomy, University of California, Irvine, California 92697, USA}
\affiliation{Stanford Institute for Materials and Energy Sciences, SLAC National Accelerator Laboratory and Stanford University, Menlo Park, California 94025, USA}
\author{Steven R. White}
\affiliation{Department of Physics and Astronomy, University of California, Irvine, California 92697, USA}
\author{A. L. Chernyshev}
\affiliation{Department of Physics and Astronomy, University of California, Irvine, California 92697, USA}
\date{\today}
\begin{abstract}
The phase diagram of the $S\!=\!1/2$ easy-axis triangular-lattice $J_1$--$J_2$ model is investigated using the density-matrix renormalization group and analytical insights. We find a significant spin-liquid  region extending from the Heisenberg limit and residing  between the Y phase---known as the magnetic analogue of the ``supersolid"---and  collinear stripe phase. The   order parameters of the supersolid are analyzed and an understanding of its lack of ferromagnetic moment  is suggested.
\end{abstract}
\maketitle
\end{CJK*}
Triangular-lattice (TL) antiferromagnets---the cradle of the spin-liquid (SL) paradigm~\cite{Balents2010QSL,Savary2016QSL,Knolle2019QSL} and an icon of  geometric frustration---continue to surprise. The current surge of interest is ignited by a new generation of the rare-earth and  transition-metal compounds, exhibiting pronounced quantum fluctuations~\cite{Haravifard2023EasyAxisTL,Sherman2023J1J2,Li2015,Tennant2024, Tennant2024KYb, Tennant2024NaYb,  Zheludev2024EasyAxisTL, Broholm2024EasyAxisTL,Nevidomskyy2024,
Prelovsek2024,
Cava2020,Li2020EasyAxis, Gao2022EasyAxisTL,Sheng2022EasyAxisTL,Gao2024EasyAxisTL, Xiang2024EasyAxisTL,Matsuda2016, Cava2019Cobaltate, Coldea2020, Christianson2024}. Our work is  inspired by  recent studies in the rare-earth compounds~\cite{Tennant2024, Tennant2024KYb, Tennant2024NaYb,Haravifard2023EasyAxisTL,Sherman2023J1J2,Li2015}, where  SL phenomenology and a broad continuum of spin excitations are seen, and  by the Ising-like  magnets~\cite{Zheludev2024EasyAxisTL, Broholm2024EasyAxisTL,Nevidomskyy2024,Prelovsek2024}, which exhibit unexpectedly strong quantum effects. A close description of many of these materials is given by the effective  $S\!=\!1/2$, $J_1$--$J_2$  $XXZ$ model
\vskip -0.1cm
\noindent
\begin{align}
\mathcal{\hat{H}} &\!=\! \sum_{n=1,2}\sum_{\langle ij \rangle_n}J_{n}\Big(  S_i^xS_j^x+S_i^yS_j^y+\Delta S_i^zS_j^z \Big),
\label{eq:spin_hamiltonian}
\end{align}
\vskip -0.2cm
\noindent
with  $\langle ij \rangle_{1(2)}$  being (next-)nearest-neighbor (NN) bonds. 

One of the earliest works on the simpler, $J_1$-only TL antiferromagnets~\cite{Fazekas1974}  sought to find an SL state near the magnetically-disordered Ising limit of the model~\cite{Wannier1950}. Later studies of the $J_1$-only easy-axis model  uncovered its unusual richness involving  order-by-disorder effects~\cite{Miyashita1985,Miyashita1986,Fazekas1992XXZ, Henley1992}. The supersolidity of the Y-phase in it has attracted significant interest  more recently from the hard-core boson perspective~\cite{Troyer2005Supersolid, Boninsegni2005Supersolid, Melko2005Supersolid, Heidarian2005Supersolid, Wang2009Supersolid, Jiang2009Supersolid, Heidarian2010Supersolid}.

\begin{figure}[t]
\includegraphics[width=\linewidth]{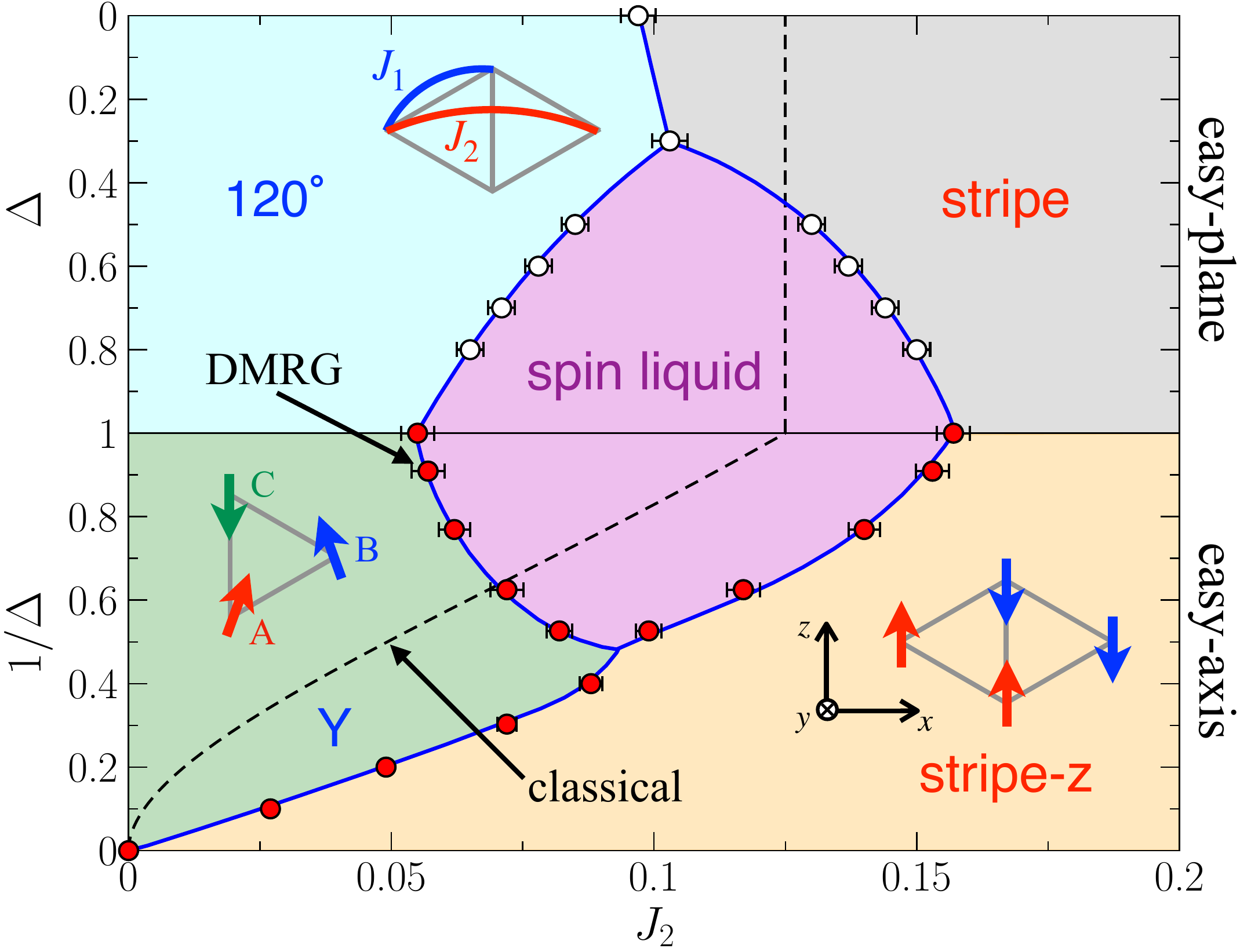}
\vskip -0.35cm
\caption{The phase diagrams for the easy-plane and easy-axis versions of the model~\eqref{eq:spin_hamiltonian}, upper panel adapted from Ref.~\cite{Zhu2017YMGO}. The solid lines are  phase boundaries interpolating transition points (symbols) from  DMRG; see text. The dashed lines are the classical phase boundaries between 120$\degree$ (Y) and stripe phases. $J_1$ and $J_2$ bonds and sketches of the Y and stripe-z states are shown; spins are in the $xz$ plane.}
\label{Fig:DMRGPHD}
\vskip -0.65cm
\end{figure}

The $J_1$--$J_2$ model (\ref{eq:spin_hamiltonian}) has been extensively studied in the isotropic Heisenberg limit, $\Delta\!=\!1$, where the presence of an SL  has been corroborated by several techniques~\cite{Zhu2015J1J2, Iqbal2016J1J2, DSheng2015J1J2,McCulloch2016J1J2, FerrariBecca2019J1J2,Pollmann2023J1J2,WietekLauchli2024QEDTL, Sheng2017,Eggert2019, Oitmaa2020J1J2, Jiang2023J1J2}. For the easy-plane model, $\Delta\!<\!1$,  the SL has been found to extend down to $\Delta\!\approx\!0.3$~\cite{Zhu2017YMGO,Iaconis2018VMC}, see Fig.~\ref{Fig:DMRGPHD}. 

However, the phase diagram of the {\it easy-axis} regime, $\Delta\!>\!1$, has been virtually unknown. There has been  no understanding of the effects of the easy-axis anisotropy on the isotropic $J_1$--$J_2$ SL state, nor has the supersolid state been sufficiently explored to meet the renewed challenges from the new compounds.

In this Letter, we  use the density-matrix renormalization group (DMRG) to explore the ground state phase diagram of the  $S\!=\!\frac{1}{2}$  easy-axis model~\eqref{eq:spin_hamiltonian} for $\Delta\!>\!1$; $J_{1}$ is set to 1. We establish the extent of the SL  phase located between the supersolid Y  and collinear ``stripe-z" phases, provide  insights into its properties, and study the complex order parameters of the ordered phases.

{\it  Phase diagram and DMRG.}---%
The phase diagram of the model (\ref{eq:spin_hamiltonian}) is shown in Fig.~\ref{Fig:DMRGPHD}. Using $1/\Delta$ in the lower panel,  we map the entire easy-axis anisotropy range from the  Ising to the Heisenberg limits to the $[0,1]$ interval. The dashed lines are classical phase boundaries (see  End Matter (EM)~\footnote{PRL has institutionalized End Matter in a recent Editorial~\cite{EndMatter}.}),  while solid lines interpolate transition points obtained by DMRG. The upper panel for the easy-plane anisotropy is shown for completeness~\cite{Zhu2017YMGO}.

The main result in Fig.~\ref{Fig:DMRGPHD} is a surprising resilience of the SL state to the symmetry-breaking, easy-axis anisotropy terms, with the resultant SL area extending to rather large $\Delta$  values. The other result is a significant {\it qualitative} deviation of the Y-to-stripe-z boundary from its classical prediction, especially near the Ising limit.

\begin{figure}[t]
\includegraphics[width=\linewidth]{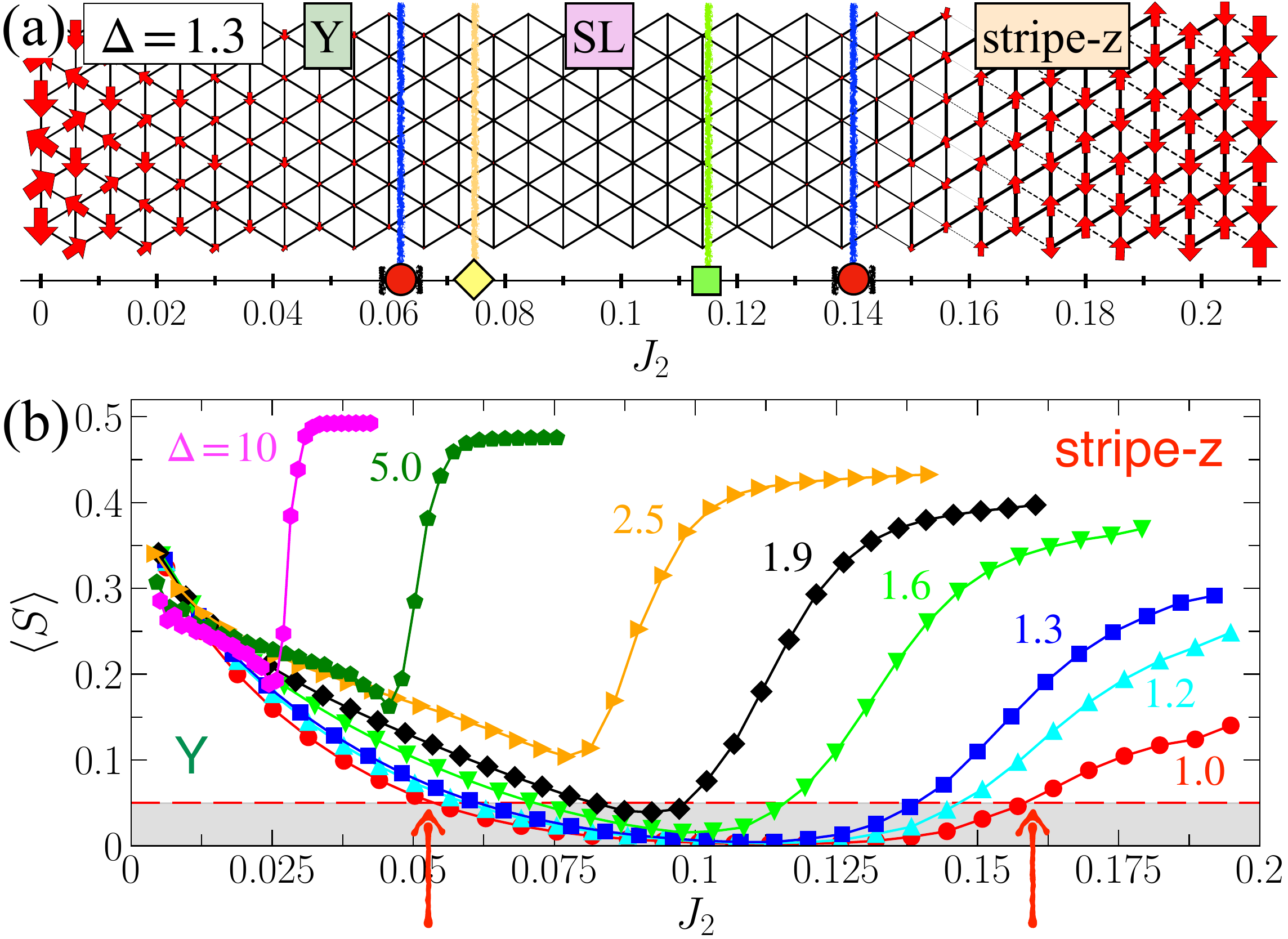}
\vskip -0.3cm
\caption{(a) $36\!\times\! 6$ DMRG scan for $\Delta\!=\!1.3$. Arrows are  ordered moments $\langle {\bf S}_i \rangle$ in the $xz$ plane and solid (dashed) lines are negative (positive)  NN correlators $\langle {\bf S}_i {\bf S}_j\rangle$. Phase boundaries from the scans (non-scans) are shown by circles (diamond and square); see the text. (b) $\langle S\rangle\!=\!|\langle {\bf S}\rangle|$ vs $J_2$ for several $\Delta$'s. The dashed line is the $\langle S\rangle\!=\!0.05$ cutoff  for the SL state.}
\label{Fig:DMRGScan}
\vskip -0.6cm
\end{figure}

We utilize several complementary approaches using DMRG simulations with the ITensor library~\cite{ITensor}. We employ  both ``scans'' (with a Hamiltonian parameter varied along the $x$-axis) and  ``non-scans'' (all parameters fixed) on the triangular-lattice $L_x\!\times\!L_y$-site open Y-cylinders (YC)~\cite{Zhu2015J1J2}, in which one of the nearest-neighbor bonds is vertical; see Fig.~\ref{Fig:DMRGScan}(a).  The spontaneous symmetry breaking in our DMRG simulations~\cite{tt'j} allows us to measure the local ordered moment $\langle {\bf S}_i\rangle$ directly, instead of, or in addition to, the spin-spin correlations. Typically, a bond dimension of $m\sim 2500$ was sufficient to ensure good convergence with the truncation error of $\mathcal{O}(10^{-6})$ when $S^z$ is not conserved. We also perform simulations with conserved $S^z$, keeping up to 8000 states and allowing the study of the wider  cylinders up to $L_y\!=\!9$.

The scans with varied $J_2$  exhibit the phases encountered along the horizontal 1D cuts of the phase diagram in Fig.~\ref{Fig:DMRGPHD}, while the non-scans on the cylinders of several sizes  provide a variety of more precise measurements. Their combination has been successfully used  in various other models and lattices~\cite{Zhu2015J1J2,Zhu2017YMGO,Zhu2018Topography,Zhu2013J1J2XYHoneycomb, Jiang2023J1J3Honeycomb,nematic2023}.

One such  $36\!\times\! 6$-cylinder scan is shown in Fig.~\ref{Fig:DMRGScan}(a) for a  representative $\Delta\!=\!1.3$. The Y and  stripe-z  phases  are clearly observed at  smaller and larger $J_2$, respectively, with their sketches shown in Fig.~\ref{Fig:DMRGPHD}. The  three-sublattice Y state has one spin along the $z$ axis and two tilted away from it by  opposite angles. It is a classical solution of the $J_1$-only easy-axis model~\cite{Miyashita1985}, which possesses additional accidental degeneracies~\cite{Miyashita1986}, and it is the groundstate of the quantum model~\cite{Henley1992,Fazekas1992XXZ}.  The stripe-z state is formed by  rows of ferromagnetic spins aligned antiferromagnetically, all pointing along the $z$ axis. 

These classical orders, with the tilt angle in the Y phase obtained from  energy minimization for a given $\Delta$ (see  EM),  are used as the pinning fields at the boundaries  in Fig.~\ref{Fig:DMRGScan}(a), and the spin patterns produce a faithful visual extent of the phases. Importantly,  the intermediate phase with strongly suppressed magnetic moment and isotropic spin-spin correlations $\langle {\bf S}_i {\bf S}_j\rangle$ on the bonds can also be observed. 

To map out the phase diagram of the model (\ref{eq:spin_hamiltonian}), we identified the phase boundaries from the plots of the ordered moment $\langle S\rangle$ averaged over the circumference of the cylinders of the  $J_2$-scans, shown in Fig.~\ref{Fig:DMRGScan}(b) for several $\Delta$'s. We adopt a criterion that $\langle S\rangle\!<\!0.05$ implies an SL state~\cite{Zhu2017YMGO}. This cutoff uses the Heisenberg limit, for which the SL boundaries  are well-settled, $0.05\!\alt\!J_2\!\alt\!0.16$~\cite{Zhu2015J1J2, DSheng2015J1J2, Iqbal2016J1J2, McCulloch2016J1J2, Oitmaa2020J1J2, Jiang2023J1J2}, and the $\Delta\!=\!1.0$ curve in Fig.~\ref{Fig:DMRGScan}(b), with the SL boundaries marked by red arrows. For $\Delta\!\agt\! 2$,  the direct Y-to-stripe-z transition is determined from the inflection point in $\langle S\rangle$ vs $J_2$. The error bars for the transitions are the $J_2$-steps in our scans.

To study the structure of the SL and ordered states, we have  employed the DMRG non-scans, discussed next.
 
\begin{figure}[t]
\includegraphics[width=\linewidth]{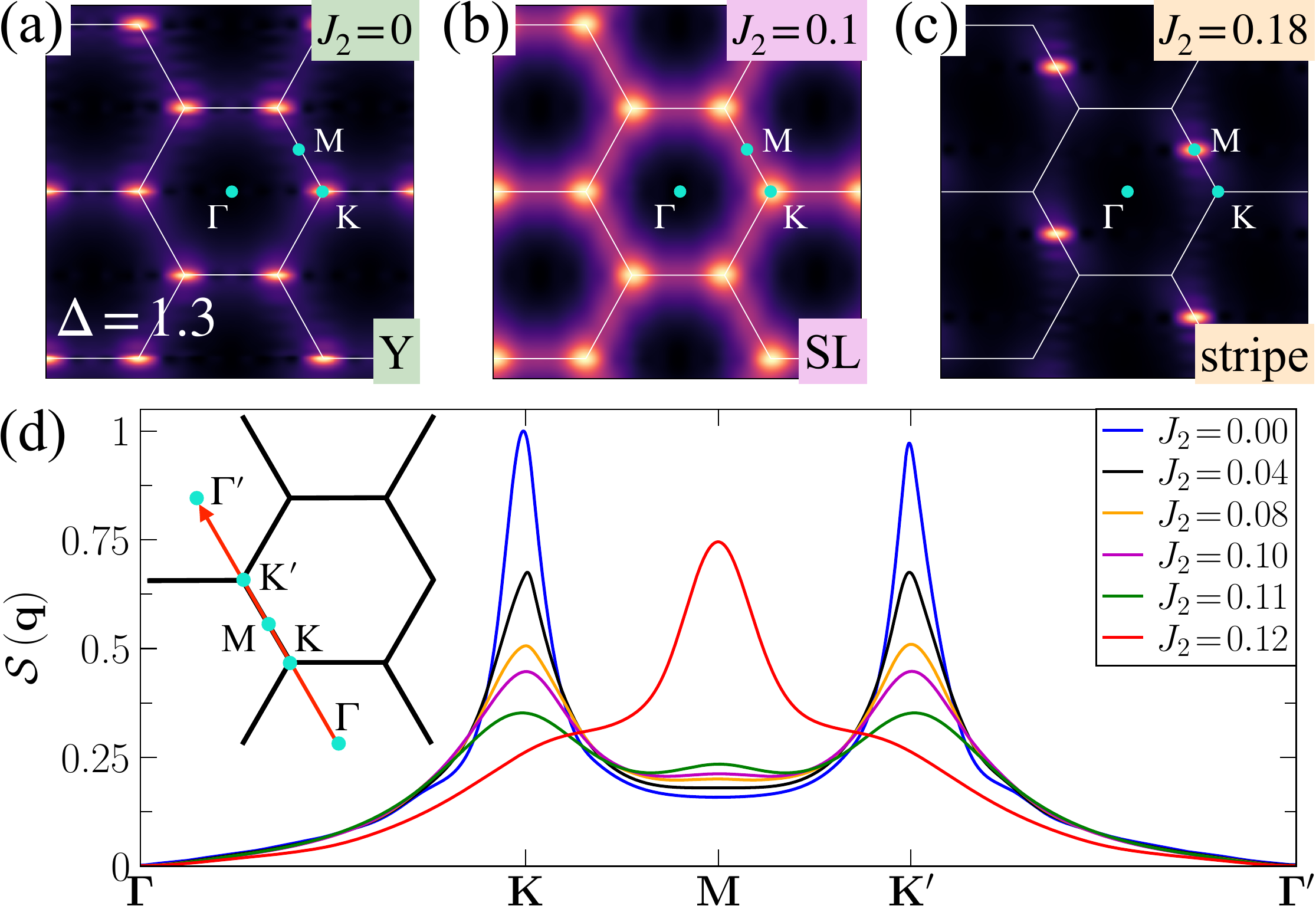}
\vskip -0.3cm
\caption{Intensity maps of ${\cal S}({\bf q})$ in the $20\!\times \!6$ YC cylinders for $\Delta\!=\!1.3$ in the (a) Y [$J_2\!=\!0$], (b) SL [$J_2\!=\!0.1$], and  (c) stripe-z [$J_2\!=\!0.18$] phases. (d)  The ${\cal S}({\bf q})$ plots  for $\Delta\!=\!1.3$ and several $J_2$ along the ${\bf q}$-path shown in the inset.}
\label{Fig:SLPLots}
\vskip -0.6cm
\end{figure}

{\it Spin-liquid and ordered phases.}---%
Our Figures~\ref{Fig:SLPLots}(a)-(c) show the intensity plots of the structure factor, ${\cal S}({\bf q})\! = \!\frac{1}{N}\sum_{i,j} e^{i{\bf q r}_{ij}} \langle {\bf S}_i {\bf S}_j\rangle$, obtained from the $20\times 6$ YC cylinders with no pinning fields~\footnote{Independence of the results from the initial DMRG state was ensured.} for $\Delta\!=\!1.3$  and $J_2$ corresponding to the (a) Y  [$J_2\!=\!0$], (b) SL [$J_2\!=\!0.1$], and (c) stripe-z  [$J_2\!=\!0.18$] states, respectively, see Fig.~\ref{Fig:DMRGPHD}.

For the Y state in Fig.~\ref{Fig:SLPLots}(a), ${\cal S}({\bf q})$ shows sharp peaks at the K-points, characteristic of the three-sublattice order, while stripe order yields peaks at  the  M-points in Fig.~\ref{Fig:SLPLots}(c), with DMRG picking one of the stripe domains. 

The structure of ${\cal S}({\bf q})$ in the SL state, Fig.~\ref{Fig:SLPLots}(b), is dominated by the broadened maxima at the K-points, allowing one to characterize it as a ``molten 120${\degree}$'' state~\cite{Maksimov2019PRX}, underscoring its  close isomorphism~\cite{Zhu2018Topography} to the isotropic case~\cite{Zhu2015J1J2}, and demonstrating the ubiquitous resilience of this SL  to all forms of anisotropy~\cite{Zhu2018Topography}. This behavior is  not inconsistent with the recently advocated scenario of the $U(1)$ Dirac SL in the isotropic $J_1$--$J_2$ model~\cite{WietekLauchli2024QEDTL,PoilblancDirac24}.

Fig.~\ref{Fig:SLPLots}(d) shows ${\cal S}({\bf q})$ for several $J_2$, normalized by the $J_2\!=\!0$ maximum at ${\bf q}$=K. The abrupt emergence of the M-point peak at $J_2\!=\!0.12$  suggests a first-order SL-to-stripe transition, while  the smooth $J_2$-evolution of the K-point peaks argues for the continuous SL-to-Y transition,  consistent with the isotropic case~\cite{Zhu2015J1J2,Iqbal2016J1J2, FerrariBecca2019J1J2,Pollmann2023J1J2,WietekLauchli2024QEDTL}.

The M-point peak in ${\cal S}({\bf q})$  indicates the SL-to-stripe transition at a smaller $J_2$ than given by the scans; see the square in Fig.~\ref{Fig:DMRGScan}(a). For the Y state,  the $1/L$-scaling of $\langle S\rangle_L$ from the middle of the clusters of fixed aspect ratio with the edges pinned by the classical Y phase~\cite{Zhu2017YMGO, Zhu2018Topography} was proposed to yield the correct 2D  limit $\langle S\rangle_\infty$~\cite{White2007HAFTL}. The $\langle S\rangle_\infty$ vs $J_2$  suggests a continuous Y-to-SL transition at a larger $J_2$ than  the scan; see the diamond in Fig.~\ref{Fig:DMRGScan}(a) and EM. These more conservative estimates still point to a rather substantial region of the SL  phase in the  phase diagram of the easy-axis model~(\ref{eq:spin_hamiltonian});  see EM.

We have performed further DMRG characterizations of the SL state using various non-scans, with and without conserving $S^z$, in the wider  cylinders up to $L_y\!=\!9$, and up to higher bond dimensions, analyzing spin-spin correlations and responses to the induced orders. We have  verified that the spin-spin correlations in the SL state show no evidence of the valence-bond or scalar chiral order, ${\bf S}_i\cdot({\bf S}_j\!\times\!{\bf S}_k)$, and that such orders decay exponentially at short distances if artificially induced in the clusters. The ordered moment also decays away from an edge pinned with the classical Y state with the correlation length $\alt\!2a$, consistent with  the analysis of the SLs in the other TL models~\cite{Zhu2015J1J2, Zhu2017YMGO, Zhu2018Topography}. The SL spin-spin correlations are demonstrated to likely retain the $SU(2)$ symmetry  away from the Heisenberg limit, despite the easy-axis character of the model (\ref{eq:spin_hamiltonian})~\cite{SS_81} and allowed symmetry breaking in DMRG;  see EM and Supplemental Material (SM)~\cite{Supplemental} for details.

\begin{figure}[t]
\includegraphics[width=\linewidth]{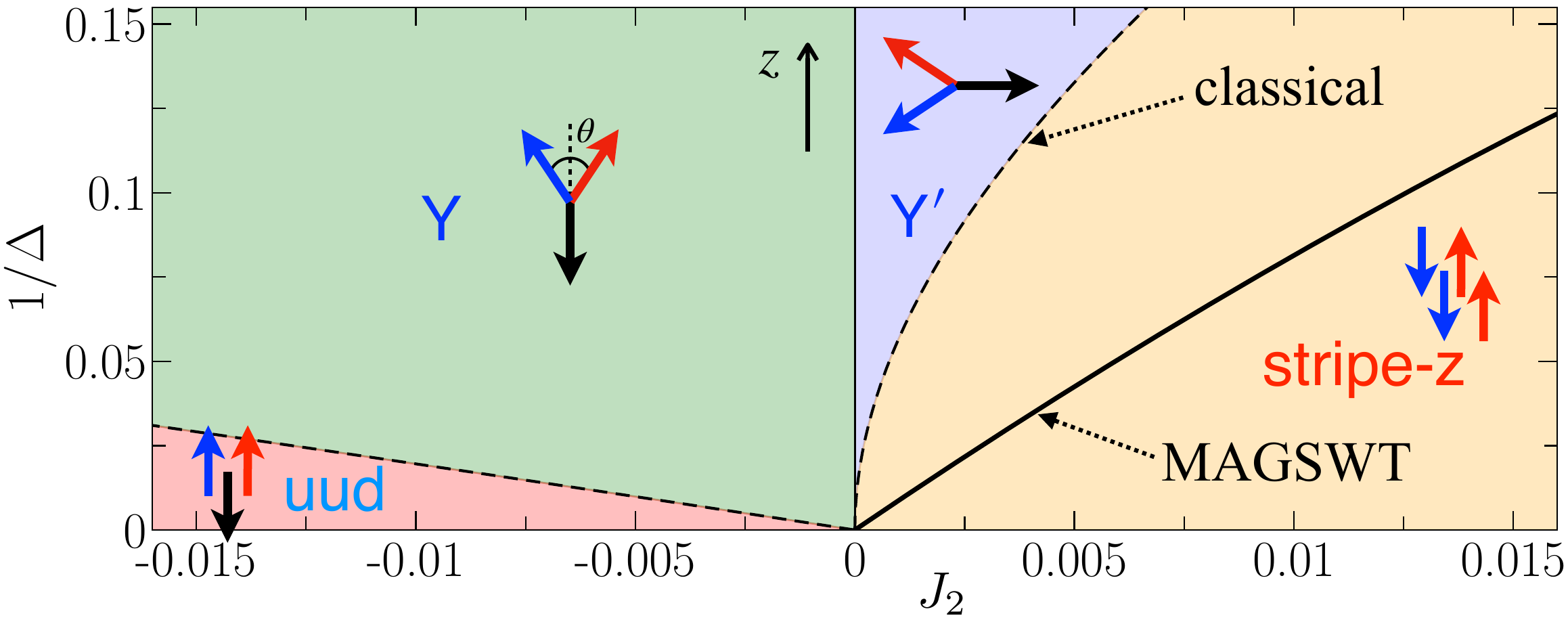}
\vskip -0.3cm
\caption{Classical phase diagram of the model (\ref{eq:spin_hamiltonian}) near the Ising point with the sketches of the states. Solid black line is the Y-to-stripe-z boundary by MAGSWT, see the text.}
\label{Fig:Quasiclassics}
\vskip -0.6cm
\end{figure}

{\it Quasiclassical analysis.}---%
Fig.~\ref{Fig:DMRGPHD} shows a notable qualitative contrast between DMRG and classical boundaries for the  Y-to-stripe-z  transition near the Ising limit. The exponential degeneracy of the Ising ground state~\cite{Wannier1950} and the associated vanishing cost of rotating a spin on each triangle in this limit  yield the classical energy of the Y-state  $E_{cl}^\mathrm{Y}\!\approx\!-1\!-\!{\bar{\Delta}}^2$ (per $NS^2J_{1}\Delta $), where $\bar{\Delta}\!=\!1/\Delta$~\cite{Fazekas1992XXZ}. It is missing the linear-$\bar{\Delta}$ term, leading to the anomalous shape of the classical phase boundary, $J_2\!\propto\!{\bar{\Delta}}^{2}$.

However,  fluctuations in the quantum case should restore the proper energy expansion near the Ising limit~\cite{Fazekas1992XXZ, Kleine1992} to produce a linear  $J_2\!\propto\!{\bar{\Delta}}$ phase boundary, as is seen in DMRG results in Fig.~\ref{Fig:DMRGPHD}. In practice, the situation is more complicated because  the {\it classical} ground state for $J_2\!>\!0$ is {\it not} a Y state, but a related  ${\rm Y}'$ state from the accidental degeneracy manifold  of the model (\ref{eq:spin_hamiltonian}) for $J_2\!=\!0$~\cite{Henley1992,Miyashita1986,Fazekas1992XXZ}, which maintains the same shape but has the reference axis in the plane instead of $z$ axis, see Fig.~\ref{Fig:Quasiclassics}. 

While  quantum fluctuations select the Y state as the ground state for $J_2\!=\!0$~\cite{Henley1992,Fazekas1992XXZ}, the same calculations cannot be performed for  $J_2\!>\!0$ as the excitation spectrum for the Y state is unstable. This conundrum is resolved  by utilizing the minimally augmented spin-wave theory (MAGSWT), which involves stabilizing spectra with  a chemical potential~\cite{Wenzel2012MAGSWT, Colleta2013MAGSWT, Jiang2023J1J3Honeycomb}. This theory supports the Y-state as the ground state  for $S\!=\!1/2$ in the entire relevant $J_2\!>\!0$ region and produces the expected  linear  Y-to-stripe-z phase boundary near the Ising point, see Fig.~\ref{Fig:Quasiclassics}, also in a broad agreement with DMRG~\cite{Supplemental}.

\begin{figure}[t]
\includegraphics[width=\linewidth]{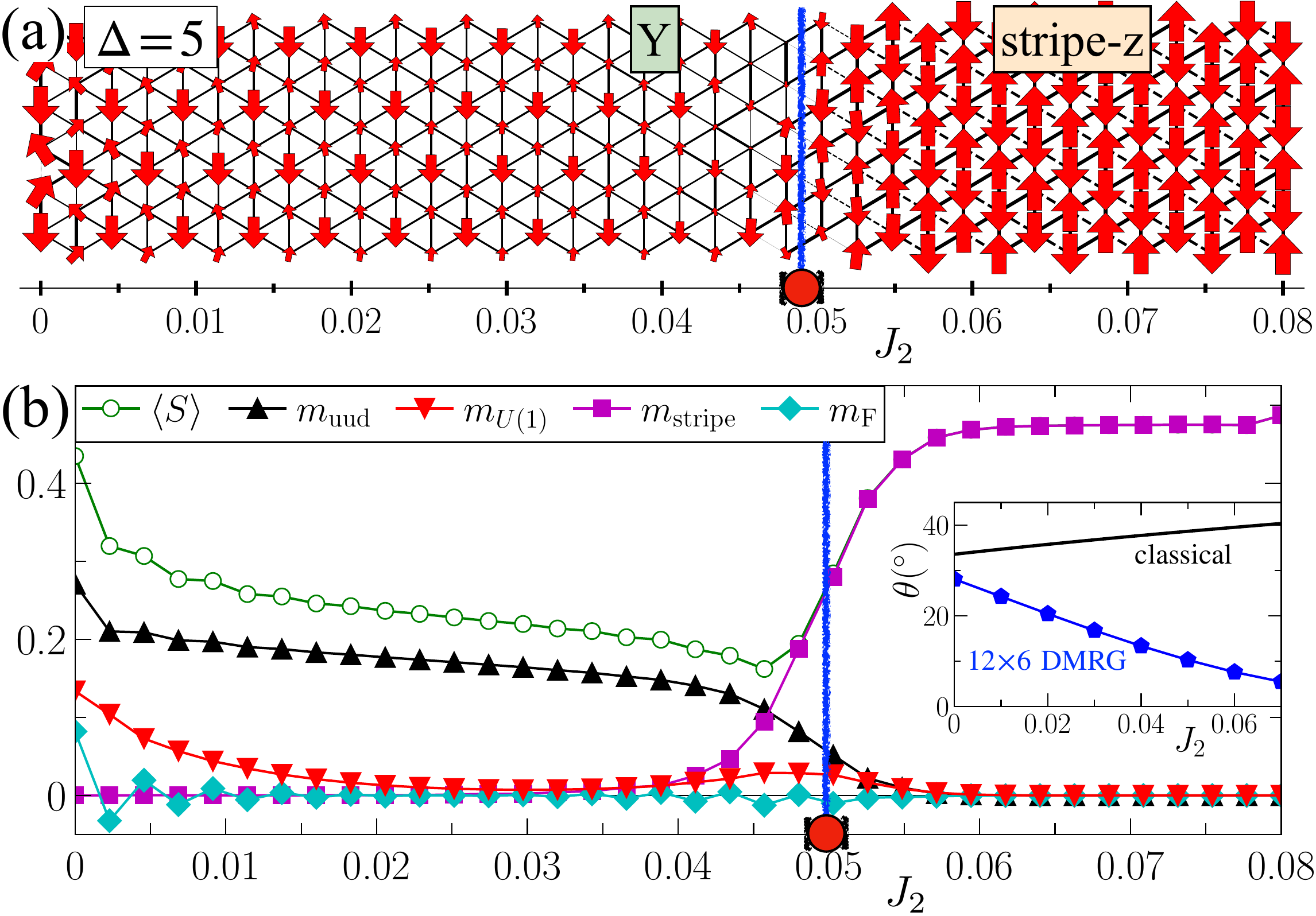}
\vskip -0.3cm
\caption{(a) Same as Fig.~\ref{Fig:DMRGScan}(a) for $\Delta\!=\!5$. (b) Order parameters vs $J_2$. Inset: tilt angle in the Y phase from the non-scans.}
\label{Fig:YtoStripe}
\vskip -0.6cm
\end{figure}

{\it Supersolid order parameters.}---%
The supersolid Y phase is described by  two order parameters, associated with the in-plane and out-of-plane components of the structure factor  ${\cal S}({\bf q})$ at the ordering vector ${\bf q}$=K~\footnote{Specifically, the solid and superfluid order parameters are defined as $m_{\mathrm{uud}}^2\!=\!\mathcal{S}^{zz}(K)/N^2$ and $m_{U(1)}^2\!=\!\mathcal{S}^{\perp}(K)/N^2$, respectively}. The ``solidity'' is the $z$-axis three-sublattice up-up-down component, $m_{\mathrm{uud}}\!=\!\frac{1}{N}\sum_{\ell} \big|\frac{1}{2}\langle S_{\ell,{\rm A}}^z\rangle\!+\!\frac{1}{2}\langle S_{\ell,{\rm B}}^z\rangle\!-\!\langle S_{\ell,{\rm C}}^z\rangle\big|$, where $\ell$ numerates unit cells and $A,B,C$ are sublattices; see Fig.~\ref{Fig:DMRGPHD}. The superfluid $U(1)$ component of the Y-state in Fig.~\ref{Fig:DMRGPHD} is $m_{U(1)}\!=\!\frac{\sqrt{3}}{2N}\sum_{\ell}\big| \langle {\bf S}_{\ell,{\rm A}}^\perp\rangle\!-\!\langle {\bf S}_{\ell,{\rm B}}^\perp\rangle\big|$, where ${\bf S}^\perp$ is the in-plane spin projection; see Ref.~\cite{Starykh_2015}. Due to the  symmetry breaking in our DMRG calculations, spins lie in the $xz$ plane, so $\langle {\bf S}^\perp\rangle \!=\!\langle S^x\rangle$. For the stripe-z phase the order parameter is  the staggered $z$-axis magnetization $m_{\mathrm{stripe}}\!=\!\frac{1}{N}\big|\sum_i e^{i{\bf qr}_i}\langle S_i^z\rangle\big|$ with the ordering vector ${\bf q}$=M.

In Fig.~\ref{Fig:YtoStripe}(a), one can observe these orders in a representative $J_2$-scan for $\Delta\!=\!5$ from the Y to the stripe-z phase. In Fig.~\ref{Fig:YtoStripe}(b), the order parameters, averaged over the circumference of the cylinder,  are plotted  vs $J_2$ together with the total ordered moment $\langle S\rangle$ and ferromagnetic moment $m_{\mathrm{F}}\!=\!\frac{1}{N}\sum_i \langle S_i^z\rangle$. The solid order $m_\mathrm{uud}$ depends weakly on $J_2$, except for the boundary and transition regions, and the stripe order is nearly classical,  both being prominent in Fig.~\ref{Fig:YtoStripe}(b). For the narrower $J_2$-scan, the width of the transition narrows too, affirming its first-order character. The transition found from the level-crossing in the non-scans coincides with the one from the inflection points in Fig.~\ref{Fig:YtoStripe}(b) within the error bars.

\begin{figure}[t]
\includegraphics[width=\linewidth]{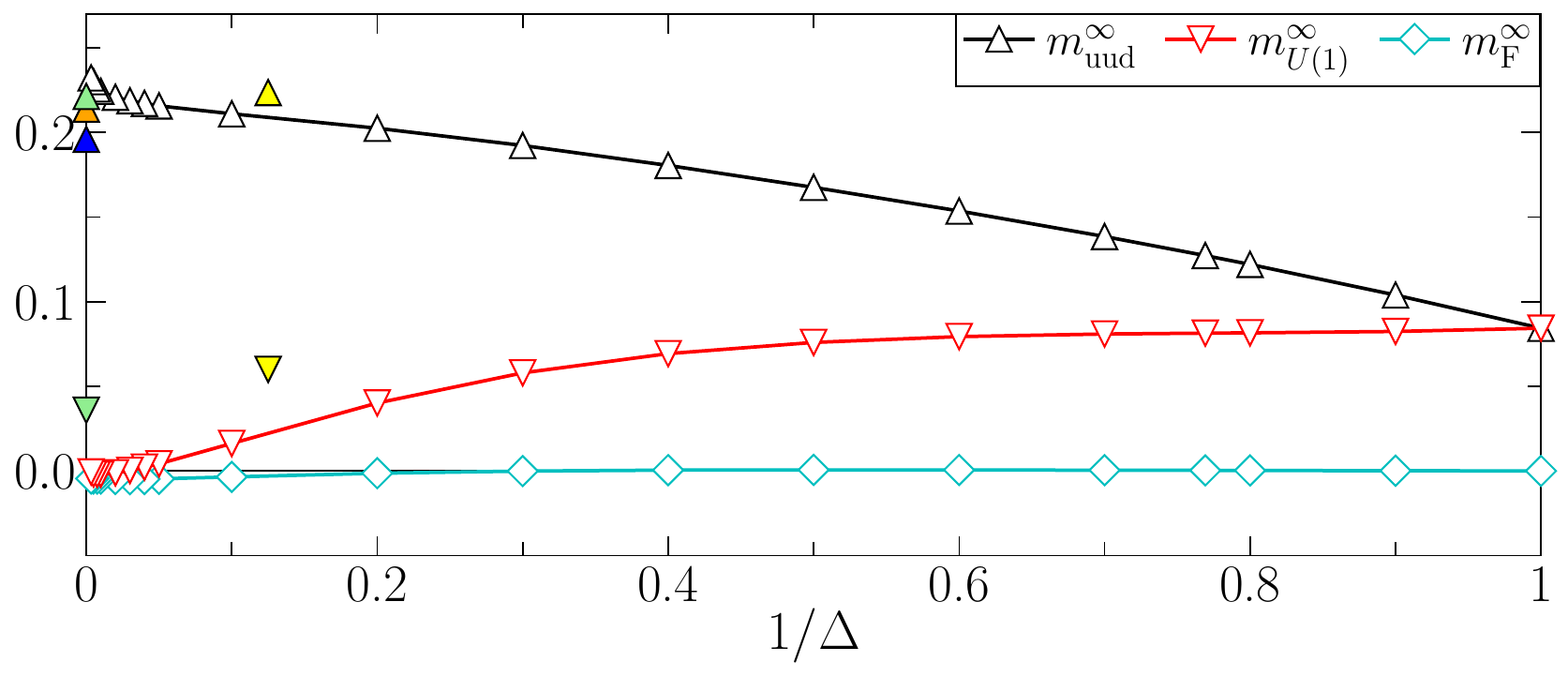}
\vskip -0.35cm
\caption{The extrapolated supersolid  and ferromagnetic order parameters vs $1/\Delta$ for $J_2\!=\!0$. Full symbols: yellow, Ref.~\cite{Jiang2009Supersolid}, DMRG; green, Ref.~\cite{Heidarian2010Supersolid}, and blue, Ref.~\cite{Wang2009Supersolid}, variational Monte Carlo (MC); orange, Ref.~\cite{Heidarian2005Supersolid}, quantum MC.}
\label{Fig:OrderParameters}
\vskip -0.65cm
\end{figure}

The superfluid order parameter $m_{U(1)}$ in Fig.~\ref{Fig:YtoStripe}(b) is more delicate. One can clearly observe small, but non-zero spin tilts in the Y-phase region in Fig.~\ref{Fig:YtoStripe}(a) all the way to the transition. The non-scans in the  $12\!\times \!6$ clusters with the classical Y-phase edges also yield a finite $m_{U(1)}$ and  tilt angle as is shown in the inset of Fig.~\ref{Fig:YtoStripe}(b). 

On the other hand, the fixed aspect ratio  $1/L$-scaling of $m_{U(1)}$, discussed above for $\langle S\rangle_\infty$ (see also EM), suggests that the extrapolated $m^\infty_{U(1)}$ vanishes somewhat before the Y-to-stripe transition.  We have tracked this behavior of $m^\infty_{U(1)}$ for several values of $\Delta$ and also  for the Y-to-SL boundary, see SM~\cite{Supplemental}. While the existence of a thin layer of the pure up-up-down state with $m_{U(1)}\!=\!0$ in the vicinity of the border to the stripe and SL phases cannot be ruled out,  possible non-linear effects in the finite-size extrapolations for the already small values of  $m_{U(1)}$ make such a scenario difficult to ascertain. In any event, at the Y-to-stripe transition, the superfluid order  $m_{U(1)}$ seem to vanish more continuously than the solid  one.

The supersolidity of the Y-phase in the $J_2\!=\!0$ version of the $S\!=\!1/2$ model~\eqref{eq:spin_hamiltonian} has attracted significant interest from the hard-core boson perspective~\cite{Troyer2005Supersolid, Boninsegni2005Supersolid, Melko2005Supersolid, Heidarian2005Supersolid, Wang2009Supersolid, Jiang2009Supersolid, Heidarian2010Supersolid}, with the major issue being whether the {\it finite} solid $m_\mathrm{uud}$  and superfluid $m_{U(1)}$ orders appear already for an infinitesimal deviation away from the Ising limit, $1/\Delta\!\rightarrow\! 0$. 

We have used the $1/L$-scaling in  fixed aspect ratio DMRG clusters  (see SM~\cite{Supplemental}) to study these order parameters for $J_2\!=\!0$. The results for the extrapolated orders, $m^\infty_{\rm uud}$ and $m^\infty_{U(1)}$, are shown in Fig.~\ref{Fig:OrderParameters} vs $1/\Delta$, together with the ferromagnetic moment $m^\infty_{\rm F}$. Also shown are the results from Refs.~\cite{Wang2009Supersolid, Jiang2009Supersolid, Heidarian2010Supersolid} obtained by various numerical methods (solid symbols).

The solid order parameter, $m_{\rm uud}$, is finite in the Ising limit, in a good quantitative agreement with the earlier studies for both small and extrapolated $1/\Delta$~\cite{Heidarian2005Supersolid, Troyer2005Supersolid, Boninsegni2005Supersolid, Melko2005Supersolid,Wang2009Supersolid, Jiang2009Supersolid, Heidarian2010Supersolid}. However, our superfluid order parameter $m_{U(1)}$ has significantly smaller values near the Ising limit than suggested previously~\cite{Jiang2009Supersolid}, and  suggests a continuous transition at $1/\Delta\!\rightarrow\! 0$, contrary to the previous works~\cite{Heidarian2010Supersolid}.

We also note that our results in Fig.~\ref{Fig:OrderParameters}, adjusted by the experimental $g$-factors, can be used for a direct comparison with the values of the supersolid order parameters in  the Ising-like  magnets~\cite{Zheludev2024EasyAxisTL, Broholm2024EasyAxisTL,Nevidomskyy2024,Cava2020,Li2020EasyAxis, Gao2022EasyAxisTL,Sheng2022EasyAxisTL,Gao2024EasyAxisTL, Xiang2024EasyAxisTL}.

{\it Ferromagnetic ordered moment.}---%
By construction, the classical Y-state has a finite ferromagnetic moment on each triangle for any $\Delta\!>\!1$~\cite{Miyashita1985}, reaching $m_{\mathrm{F}}\!=$1/3  in the Ising limit.  For the $S\!=\!1/2$ case, Refs.~\cite{Kleine1992,Fazekas1992XXZ}  showed that strong fluctuations of the tilted spins lead to a suppressed $m_{\mathrm{F}}\!\alt\!0.07$, suggesting that it vanishes. The  hard-core bosons studies  in the Ising  limit found finite, but  small values of $m_{\mathrm{F}}\!\approx\!0.01$~\cite{Jiang2009Supersolid,Wang2009Supersolid}.

As one can see in Figs.~\ref{Fig:YtoStripe}(b) and \ref{Fig:OrderParameters}, the ferromagnetic order parameter is essentially zero throughout the extent of the Y phase, aside from the boundary in Fig.~\ref{Fig:YtoStripe}(b) due to the classical pinning field.  Even near the Ising limit, where $m^\infty_{\mathrm{F}}$ deviates from zero in Fig.~\ref{Fig:OrderParameters} due to the accuracy of our approach, reaching $m^\infty_{\mathrm{F}}\!\approx\! -5 \!\times \! 10^{-3}$, it is   still one-to-two orders of magnitude smaller than the discussed small values of the $m_{U(1)}$ order parameter, giving a strong indication that $m_{\mathrm{F}}$, in fact, must be zero. 

To rationalize such a cancellation of the ordered moments  {\it on each triangle}, we rewrite the $J_1$ part of \eqref{eq:spin_hamiltonian} as
\vskip -0.15cm
\noindent
\begin{equation}
\frac{\hat{\cal H}_1}{J_1} \!= \!\frac{1}{2}\sum_{\triangle}\!\Big(\big(S_{\triangle}^\perp\big)^2\!+\!\Delta\big(S_{\triangle}^z\big)^2\Big)\!-\!\frac{3}{2}(\Delta\!-\!1)\sum_i\! \big(S_{i}^z\big)^2,
\label{eq:H_triangle}
\end{equation}
\vskip -0.25cm
\noindent
relative to $E_0\!=\!-3NS(S+1)/2$, with the first sum over the triangles of one orientation and $S^\alpha_\triangle\!=\!S^\alpha_A+S^\alpha_B+S^\alpha_C$ on such a triangle. A similar rewriting using the full squares of ${\bf S}_{\triangle}$ is common for the Heisenberg TL model~\cite{Baskaran1989}. 

The key feature of the Hamiltonian in Eq.~\eqref{eq:H_triangle}, is that the {\it second} sum is a constant  for $S\!=\!1/2$, naturally suggesting the ground state that resides in the $S^\alpha_{\triangle}\!=\!0$ sector on each triangle and minimizes $(S_\triangle^\alpha)^2$ for each spin component for all values of  anisotropy $\Delta$.  This offers an intuitive  explanation of the absence of the ferromagnetic order in the $S\!=\!1/2$ TL system. 

{\it Summary.}---%
We have investigated the phase diagram of the  $S\!=\!1/2$ easy-axis $J_1$--$J_2$ TL model using DMRG and analytical insights. We found significant evidence of an SL state that is surprisingly resilient to the symmetry-breaking anisotropies and presented a detailed description of it. Our findings provide the necessary framework and important theoretical guidance to the ongoing searches of the SL and other exotic states in the rare-earth and transition-metal TL compounds~\cite{Haravifard2023EasyAxisTL,Sherman2023J1J2,Li2015,Tennant2024, Tennant2024KYb, Tennant2024NaYb, Matsuda2016, Cava2019Cobaltate, Coldea2020, Christianson2024}. 

The strongly-fluctuating supersolid Y phase has also been analyzed, with the quantitative characterization of its order parameters bearing on the recent  experimental studies of the  Ising-like  magnets~\cite{Zheludev2024EasyAxisTL, Broholm2024EasyAxisTL,Nevidomskyy2024,Cava2020,Li2020EasyAxis, Gao2022EasyAxisTL,Sheng2022EasyAxisTL,Gao2024EasyAxisTL, Xiang2024EasyAxisTL}. A physical understanding of the enigmatic absence of the ferromagnetic component of the order parameter in this phase  that is also unique to the $S\!=\!1/2$ systems is offered.

\begin{acknowledgments}
{\it Acknowledgments.}---%
We would like to thank Federico Becca and Roderich Moessner for inspirational conversations. The work of  C.~A.~G. and A.~L.~C. on the numerical and analytical phase diagrams and other results was supported by the U.S. Department of Energy, Office of Science, Basic Energy Sciences under Award No. DE-SC0021221. C.~A.~G. acknowledges support by the Graduate Fellowship from Eddleman Quantum Institute at UC Irvine. The work by  S.~J. and S.~R.~W. was supported by the National Science Foundation under DMR-2412638. 
S.~J. is also supported by the Department of Energy (DOE), Office of Sciences, Basic Energy Sciences, Materials Sciences and Engineering Division, under Contract No. DEAC02-76SF00515.
We would like to thank Aspen Center for Physics and the Kavli Institute for Theoretical Physics (KITP) where different stages of this work were advanced. The Aspen Center for Physics is supported by National Science Foundation Grant No. PHY-2210452 and KITP is supported by the National Science Foundation under Grant No. NSF PHY-2309135.
\end{acknowledgments}

\bibliography{J1J2XXZTL}

\onecolumngrid
\begin{center}
\ \vskip 0.2cm
{\large\bf End Matter}
\end{center}
\twocolumngrid

\renewcommand{\theequation}{A\arabic{equation}}
\setcounter{equation}{0}

{\it Classical phase diagram.}---%
The classical energies of the Y and Y$^\prime$ states as a function of the canting angle $\theta$ measured from the easy-axis, see Fig.~\ref{Fig:Quasiclassics}, are
\begin{eqnarray}
E^\mathrm{Y}_{cl}\!=\! 1\!-\!2 \cos \theta\!+\!3J_2\!-\!(1+\bar{\Delta} +2J_2(1-\bar{\Delta}))\!\sin^2\!\theta,\quad\quad \label{eqEM:EclYtheta} \\
E^{\mathrm{Y}'}_{cl}\!\!=\!(1\!-\!2\sin \theta\!+\!3J_2)\bar{\Delta}  \!-\!(1\!+\!\bar{\Delta} \!-\!2J_2( 1\!-\!\bar{\Delta}))\!\cos^2\!\theta,\quad\quad\label{eqEM:EclYptheta}
\end{eqnarray}
in units of $NS^2J_{1}\Delta$, with $N$ the number of sites, and we use $\bar{\Delta}\!=\!1/\Delta$, as before. The minimization of Eqs.~\eqref{eqEM:EclYtheta} and \eqref{eqEM:EclYptheta} yields the classical angles
\begin{align}
\cos \theta_{\mathrm{Y}} &= \big(1+\bar{\Delta} +2J_2 (1-\bar{\Delta})\big)^{-1}, \label{eqEM:thetaclY} \\ 
\sin \theta_{\mathrm{Y}'} &=\bar{\Delta} \big(1+\bar{\Delta} -2J_2(1- \bar{\Delta})\big)^{-1},
\label{eqEM:thetaclYp}
\end{align}
which are used to obtain 
\begin{align}
&E_{cl}^\mathrm{Y} = -1+J_2(1+2\bar{\Delta})-\frac{\bar{\Delta}^2-2J_2(1-\bar{\Delta})^2}{1+\bar{\Delta} +2J_2 (1-\bar{\Delta}) }, \label{eqEM:EclY} \\
&E_{cl}^{\mathrm{Y}'} =-1+J_2(\bar{\Delta}+2) -\frac{\bar{\Delta}^2}{1+\bar{\Delta} -2 J_2(1-\bar{\Delta}) }.
\label{eqEM:EclYp}
\end{align}
The classical energies of the up-up-down (UUD) and stripe-z states do not require energy minimization and are given by $E_{cl}^{\mathrm{uud}}\!=\!-1\!+\!3J_2$ and $E_{cl}^{\mathrm{stripe}}\!=\!-1\!-\!J_2$, respectively, also in units of $NS^2J_{1}\Delta$.

The classical boundaries shown in Figs.~\ref{Fig:DMRGPHD} and  \ref{Fig:Quasiclassics} were obtained using the classical energies above. The analytical expressions for them are given in SM~\cite{Supplemental}.

Expanding Eqs.~\eqref{eqEM:EclY} and \eqref{eqEM:EclYp} near the Ising limit and in small $J_2$ gives  $E_{cl}^{\mathrm{Y}}\!\approx\! -1\!-\!\bar{\Delta}^2+3J_2$ and $E_{cl}^{\mathrm{Y}'}\!\approx\! -1\!-\!\bar{\Delta}^2+2J_2$, respectively. This yields the Y-to-stripe transition line as  $\bar{\Delta}^2\!\approx\!4J_2$ and Y$'$-to-stripe transition line  as $\bar{\Delta}^2\!\approx\!3J_2$, both showing the anomalous nonlinear type of the boundary discussed in the main text. 

For the UUD-to-Y transition, the expansion leads to $E_{cl}^{\mathrm{Y}}\approx E_{cl}^{\mathrm{uud}}-(\bar{\Delta}+2J_2)^2$, yielding the ``regular,'' i.e., linear Y-to-UUD transition boundary $\bar{\Delta}\!\approx\!-2J_2$. This is because the corresponding transition is  continuous, not a level-crossing as in the Y-to-stripe case, since the Y phase can be deformed into the UUD without breaking additional symmetries.

\begin{figure}[b]
\vskip -0.3cm
\includegraphics[width=\linewidth]{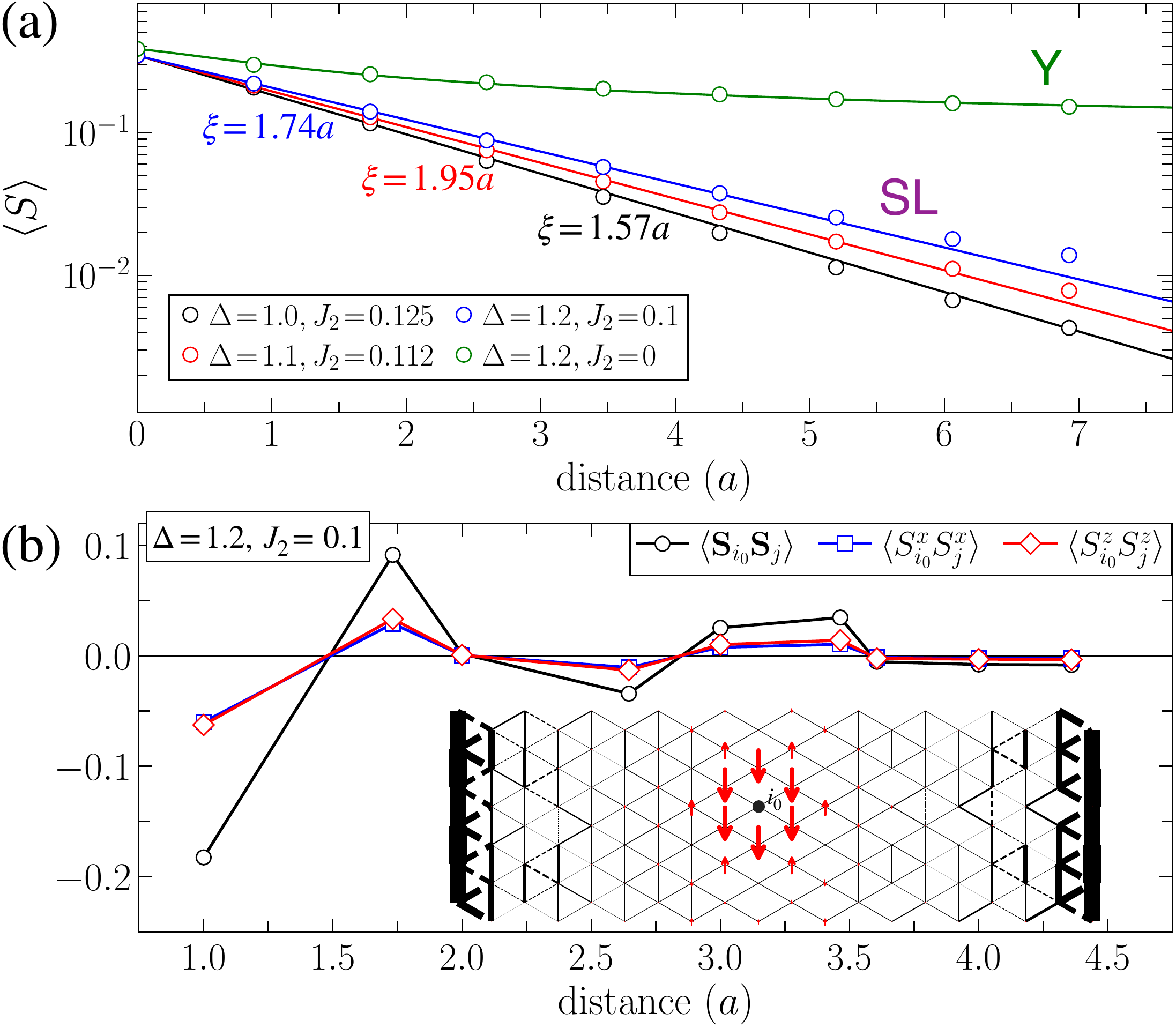}
\vskip -0.2cm
\caption{(a) $\langle S\rangle$ vs the  distance from the edge in the $20\times6$ YC cylinder for the parameters  from the SL region marked as stars in Fig.~\ref{FigEM:NonScanPHD}(a) and with the classical Y state pinning on the left edge. The solid lines are the exponential fits with the correlation length $\xi$. (b) The real-space spin-spin correlations $\langle {\bf S}_{i_0} {\bf S}_j\rangle$ and their $xx$ and $zz$ components in the $20\times6$ YC cylinder in the SL region without  pinning fields. Inset: The circle marks the $i_0$ site, bonds are the NN $\langle {\bf S}_{i} {\bf S}_j\rangle$ correlators with the average subtracted, and  arrows represent the values of the $\langle {\bf S}_{i_0} {\bf S}_j\rangle$.}
\label{FigEM:SLPlots2}
\end{figure}

{\it Correlations in the SL state.}---%
For the SL state, we have  investigated the decay of the induced ordered moment away from the boundary of the $20\times 6$ YC cylinders for several representative parameter choices from the putative SL region, with the pinning field applied to one of the open ends of the cylinder in the form of the classical Y state. It was verified that this pinning is the one that exhibits the slowest decay. For the ordered state, one expects a power-law decay to a constant value. 

\begin{figure}[t]
\includegraphics[width=\linewidth]{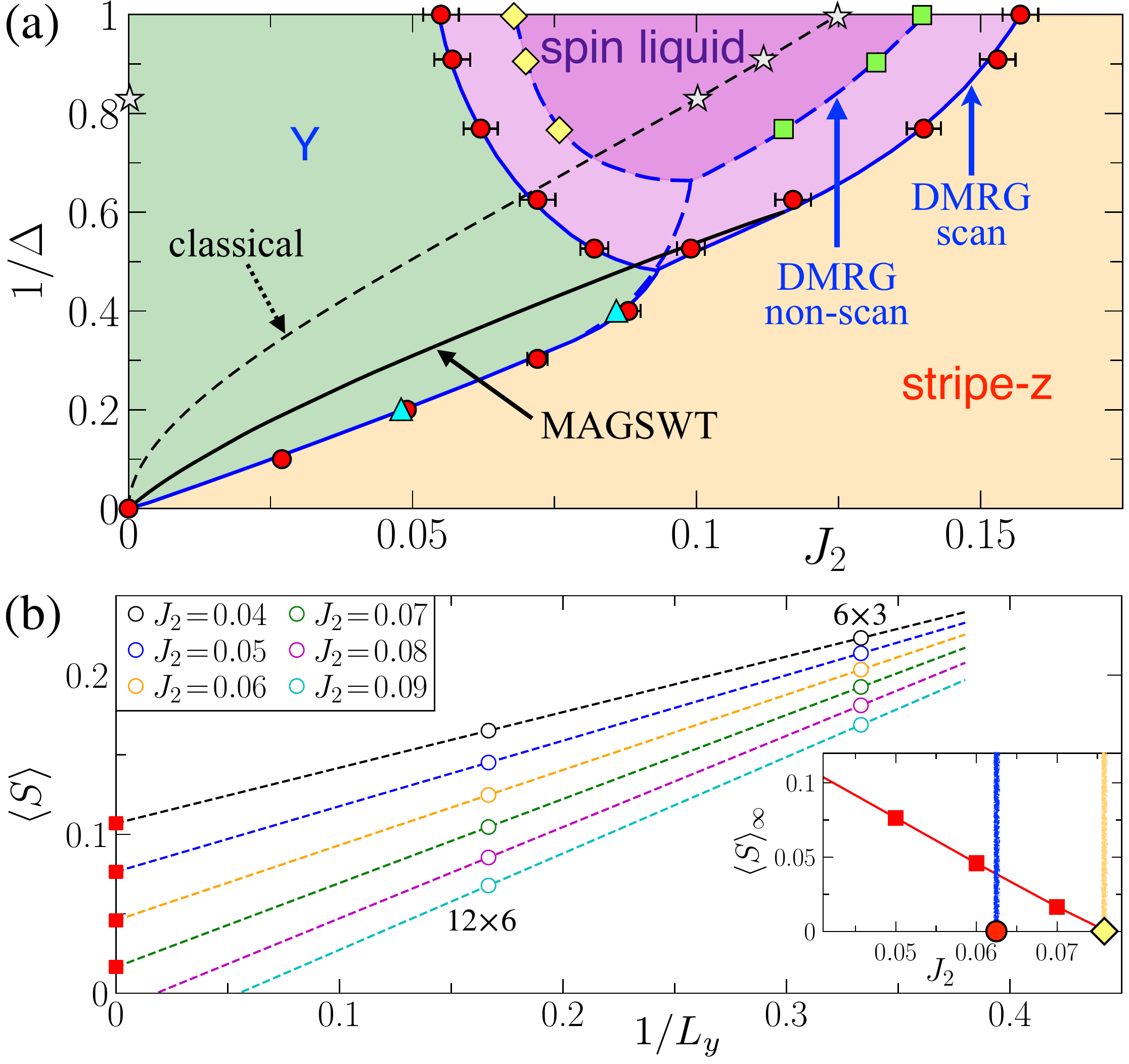}
\vskip -0.2cm
\caption{(a) Same as the easy axis panel in Fig.~\ref{Fig:DMRGPHD}. The blue dashed lines are the phase boundaries interpolating transition points (symbols) from the analyses using the DMRG non-scans. The solid black line is the MAGSWT Y-to-stripe-z boundary. Other symbols are as described in the text. (b) $1/L$-scaling of $\langle S\rangle$ for $\Delta\!=\!1.3$ and several $J_2$. Inset: extrapolated $\langle S\rangle_\infty$ vs $J_2$; the circle  and diamond are the Y-SL phase boundaries from the scan in Fig.~\ref{Fig:DMRGScan} and extrapolation, respectively.}
\label{FigEM:NonScanPHD}
\vskip -0.5cm
\end{figure}

Our Fig.~\ref{FigEM:SLPlots2}(a) shows the ordered moment $\langle S\rangle$ along the length of the cylinder for the values of $\Delta$ and $J_2$ from the SL region, marked as stars in Fig.~\ref{FigEM:NonScanPHD}(a). The semi-log plot highlights the clear exponential decay of the induced order for all three sets, with the rather short correlation lengths, $\xi\!<\!2a$, comparable to the similar analysis in Refs.~\cite{Zhu2015J1J2,Zhu2018Topography}. It is worth noting that this type of measurement is not practical for identifying  phase boundaries of the SL region, as the power-law decay of the order parameter can be hard to discern for the border region and weak orders.

We provide additional characterization of the SL state in our Fig.~\ref{FigEM:SLPlots2}(b) that shows the  decay of the real-space spin-spin correlations $\langle {\bf S}_{i_0} {\bf S}_j\rangle $ in the $20\times 6$ YC cylinder  for $\Delta\!=\!1.2$ and $J_2\!=\!0.1$  from the SL region. Similar analysis with nearly identical results has been performed in the $20\times 9$ YC cylinder with conserved $S^z$ and keeping up to 8000 states.

 The $i_0$ site  in the center of the cluster in the inset, at which $\langle {\bf S}_{i_0}^2\rangle\!=\!3/4$, is marked by the black circle. The length of the up (down) arrows on each site indicates the positive (negative) values of the correlations, also shown in the main figure. In Fig.~\ref{FigEM:SLPlots2}(b), we also show $xx$ and $zz$ components of $\langle {\bf S}_{i_0} {\bf S}_j\rangle$ separately ($yy$ is identical to $xx$) in order to demonstrate that  spin-spin correlations in the  SL state seem to retain the $SU(2)$ symmetry, despite the easy-axis character of the model (\ref{eq:spin_hamiltonian}) and  symmetry breaking allowed in DMRG. While a  systematic study of this symmetry enrichment~\cite{SS_81} is needed,   the difference of the NN $xx$ and $zz$  components is less than 5\% for $\Delta\!=\!1.2$, significantly away from the Heisenberg limit.

The thickness of the bonds in the inset in Fig.~\ref{FigEM:SLPlots2}(b) is proportional to the NN correlators $\langle {\bf S}_i {\bf S}_j\rangle $ with the subtracted average value of $-0.175$. The result of the subtraction is magnified to make the subtler variations from the average more visible; see Refs.~\cite{Zhu2015J1J2,Zhu2018Topography}. Aside from the boundary-effects, we do not observe any sign of the valence-bond orders. A more detailed verification that the spin-spin correlations in the SL state show no evidence of the responses to the artificially induced orders, such as valence-bond solid and scalar chiral order, ${\bf S}_i\cdot({\bf S}_j\!\times\!{\bf S}_k)$,  is given in SM~\cite{Supplemental}, utilizing  conserved $S^z$ non-scans with higher bond dimensions.

{\it Conservative phase diagram.}---%
Our Fig.~\ref{FigEM:NonScanPHD}(a) shows the easy-axis panel of the phase diagram from Fig.~\ref{Fig:DMRGPHD} superimposed with the additional data from the DMRG  non-scans and results from the analytical  MAGSWT approach, see the main text. 

The star symbols along the classical phase boundary from the SL region are the representative parameters studied above, see Fig.~\ref{FigEM:SLPlots2}(a). The triangles along the Y-to-stripe-z transition boundary show the  energy-level crossings of these competing states obtained from the $20\times 6$ non-scan clusters without pinning field for $\Delta\!=\!2.5$ and $5.0$~\cite{Supplemental}. The results agree closely with the ones obtained from the scans. 

The SL-to-stripe transition points, marked by the squares, are obtained from the $20\!\times\! 6$ clusters with no pinning fields for $\Delta\!=\!1.0$, $1.1$, and $1.3$, as the $J_2$ values at which the peaks at the M-point in the structure factor ${\cal S}({\bf q})$ become pronounced, see Fig.~\ref{Fig:SLPLots} and discussion in the main text. 

The Y-to-SL  transition points for the same values of $\Delta$ are marked as diamonds. They are derived from the $1/L$-extrapolation of the ordered moment, with the representative process shown in Fig.~\ref{FigEM:NonScanPHD}(b) for $\Delta\!=\!1.3$ and several $J_2$. It uses  the non-scans with fixed aspect ratio ($6\times 3$ and $12\times 6$) that can fit the three-sublattice state without frustration and extrapolates the ordered moment $\langle S\rangle$ from the middle of the clusters with the edges pinned by the classical  Y-order to the thermodynamic limit, $\langle S\rangle_\infty$. It has been argued that keeping the same aspect ratio is essential to approach such a limit consistently~\cite{White2007HAFTL}.  Then, the  $\langle S\rangle_\infty$ vs $J_2$, shown in the inset, allows us to find the Y-to-SL transition  for a given $\Delta$. The transition obtained from the scan using the criterion discussed in the main text is shown for comparison (circle).

The blue dashed line interpolates these more conservative estimates of the SL boundaries, still leaving a substantial region of the SL phase in the phase diagram of the easy-axis model (\ref{eq:spin_hamiltonian}). The  solid black line is the Y-to-stripe MAGSWT transition line, see  Fig.~\ref{Fig:Quasiclassics} and SM~\cite{Supplemental}. 

\begin{figure}[t]
\includegraphics[width=\linewidth]{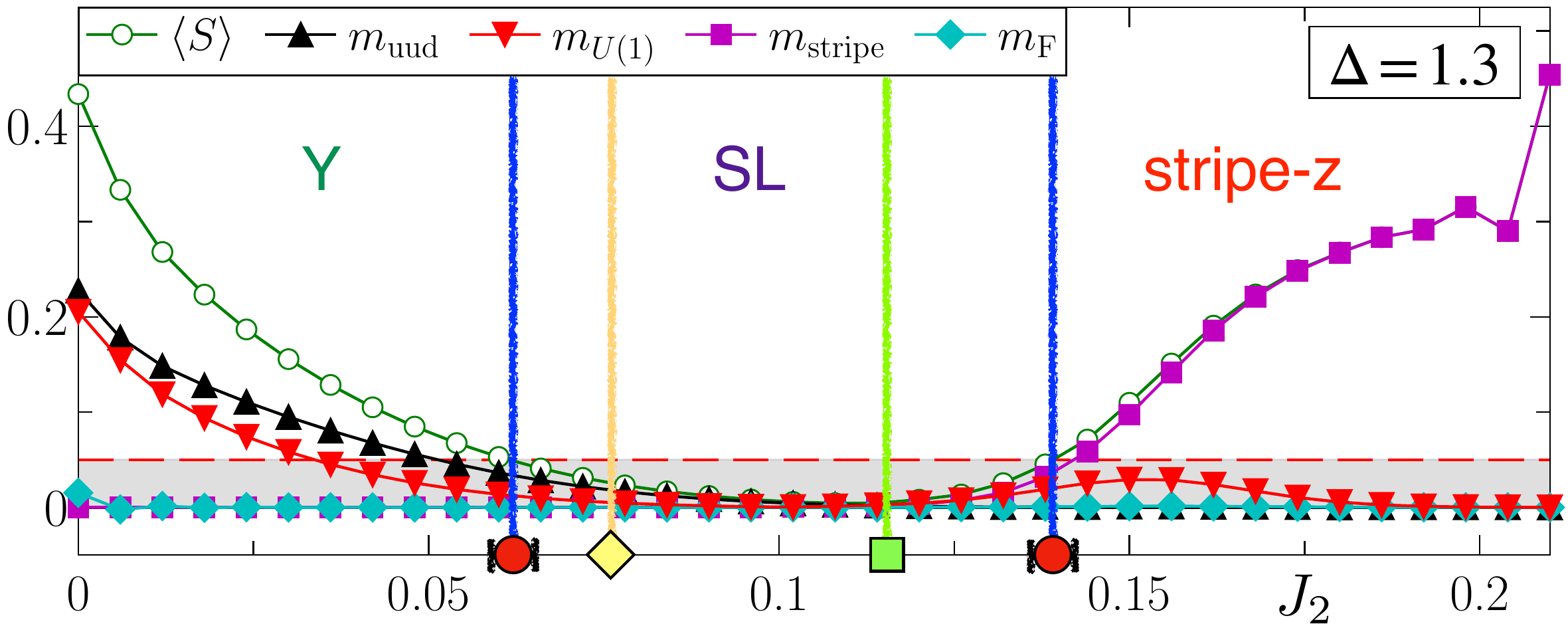}
\vskip -0.2cm
\caption{Same as in Fig.~\ref{Fig:YtoStripe}(b) for $\Delta\!=\!1.3$.}
\label{FigEM:ScanD1p3}
\end{figure}

{\it More order parameters.}---%
In Fig.~\ref{FigEM:ScanD1p3}, we show order parameters vs $J_2$ for the  $\Delta\!=\!1.3$ scan in Fig.~\ref{Fig:DMRGScan}(a). Notations are the same as in Fig.~\ref{Fig:YtoStripe}(b). This figure shows the SL boundaries according to different criteria as in Fig.~\ref{Fig:DMRGScan}(a), and the behavior of all shown quantities is consistent with the discussions in the main text. Plots like this are useful for our analysis provided in main text.



\newpage 
\ \
\newpage 
\ \
\newpage
\onecolumngrid
\begin{center}
{\large\bf Phase Diagram of the Easy-Axis Triangular-Lattice  $J_1$--$J_2$ Model: \\ Supplemental Material}\\ 
\vskip0.35cm
Cesar A. Gallegos,$^1$ Shengtao Jiang,$^{1,2}$ Steven R. White,$^1$ and A. L. Chernyshev$^1$\\
\vskip0.15cm
{\it \small $^1$Department of Physics and Astronomy, University of California, Irvine, California
92697, USA}\\
{\it \small $^2$Stanford Institute for Materials and Energy Sciences, SLAC National Accelerator Laboratory and Stanford University, \\ Menlo Park, California 94025, USA}\\
{\small (Dated: \today)}\\
\end{center}
\vskip -0.5cm \

\setcounter{page}{1}
\thispagestyle{empty}
\makeatletter
\renewcommand{\c@secnumdepth}{0}
\makeatother
\setcounter{section}{0}
\renewcommand{\theequation}{S\arabic{equation}}
\setcounter{equation}{0}
\renewcommand{\thefigure}{S\arabic{figure}}
\setcounter{figure}{0}
\renewcommand{\thetable}{S.\Roman{table}}
\setcounter{table}{0}
\vspace{-0.3cm}
\section{DMRG details}

\vspace{-0.3cm}
\subsection{Technical details}

\vspace{-0.3cm}
\subsubsection{Extrapolation with the truncation error}
\vskip -0.2cm

For most of the DMRG calculations, we systematically increase the bond dimension $m$ at every other DMRG sweep. To obtain local observables, such as the order parameters, we typically average their values at the center of the cluster and perform a linear fit as a function of the truncation error using the results from the second sweeps of the last three pairs of sweeps to extrapolate to zero truncation error~\cite{White2007HAFTL}.

\vspace{-0.3cm}
\subsubsection{Conservation of quantum numbers}
\vskip -0.2cm

\begin{figure}[h!]
\includegraphics[width=.9\linewidth]{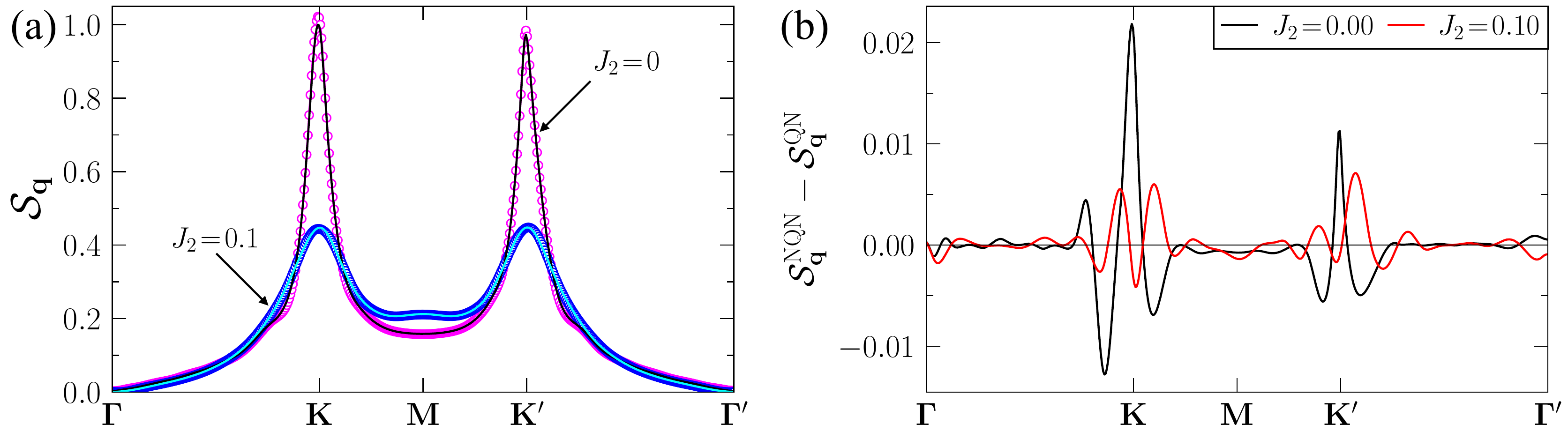}
\vskip -0.3cm
\caption{(a) Similar to Fig.~\ref{Fig:SLPLots}(d), for $\Delta\!=\!1.3$ and $J_2\!=\!0$ and $0.1$. Open symbols are the DMRG results without conserving $S^z$ and solid lines are the results for $S^z$ conserved. (b) The difference between the $\mathcal{S}({\bf q})$ results in (a).}
\label{FigSM:SqQNvsNQN}
\vskip -0.3cm
\end{figure}

Since we have established that the ground state of the model~\eqref{eq:spin_hamiltonian} lies in the $S^z_{\mathrm{total}}=0$ sector, we performed additional non-scan calculations that conserve the quantum number $S^z$ to verify the results presented in Fig.~\ref{Fig:SLPLots}. Fig.~\ref{FigSM:SqQNvsNQN}(a) shows nearly indistinguishable static structure factor between the $S^z$-conserved and non-conserved DMRG results, $\mathcal{S}_{\bf q}^{\mathrm{QN}}$ and $\mathcal{S}_{\bf q}^{\mathrm{NQN}}$, respectively, for $\Delta\!=\!1.3$ and $J_2\!=\!0$ $(0.1)$ in the Y (spin-liquid) phase. Fig.~\ref{FigSM:SqQNvsNQN}(b) shows that the numerical difference does not exceed $2\%$. Additionally, we verified that the difference in the final converged energy per site, for $m\!=\!2500$ without conserving $S^z$ and $m\!=\!5000$ with $S^z$ conserved, is of the order $\mathcal{O}(10^{-6})$, indicating that both approaches are reliably converging to the ground state.

\vspace{-0.3cm}
\subsection{Further SL analysis}

\vspace{-0.3cm}
\subsubsection{VBS checks}
\vskip -0.2cm

\begin{figure}[h!]
\includegraphics[width=\linewidth]{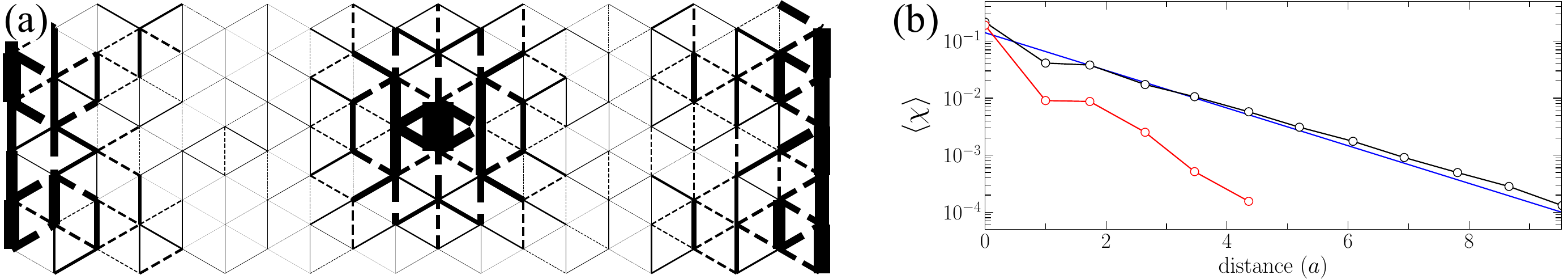}
\vskip -0.3cm
\caption{(a) The $20\!\times\!6$ non-scan YC cylinder for $\Delta\!=\!1.2$ and $J_2\!=\!0.1$ with the stronger bond in the center. Bonds are proportional to the NN correlations with the average $-0.18$ subtracted for clarity. The NN correlation $\langle {\bf S}_i {\bf S}_j\rangle$ for the pinned bond in the center is $-0.605$.
(b) The exponential decay of the chirality, $\langle\chi\rangle\!=\!\langle{\bf S}_i\cdot({\bf S}_j\!\times\!{\bf S}_k)\rangle$, in the $20\!\times\!6$ non-scan YC cylinders for $\Delta\!=\!1.2$ and $J_2\!=\!0.1$ away from the edge with the chiral pinning (black symbols) and from  the chiral pinning at the center (red symbols); see text.}
\label{FigSM:VBS_Chirality}
\vskip -0.3cm
\end{figure}

Fig.~\ref{FigSM:VBS_Chirality}(a) shows the $20\!\times \!6$ non-scan  YC cylinder for $\Delta\!=\!1.2$ and $J_2\!=\!0.1$ from the SL phase, without the pinning fields and with $S^z$ conserved. For one vertical bond in the center of the cluster, the NN correlator $\langle {\bf S}_i {\bf S}_j\rangle$ is artificially enhanced by doubling the value of the $J_1$-exchange on this bond, in order to study the response of the suggested SL state to the valence-bond solid (VBS) formation.

The thickness of the bonds in Fig.~\ref{FigSM:VBS_Chirality}(a) is proportional to the NN correlators with their average value of $-0.18$  subtracted for clarity. The result of the subtraction is magnified to make the subtler variations from the average more visible. Aside from the boundary and the immediate vicinity of the perturbed bond, we do not observe any sign of a possible VBS pattern, suggesting the absence of these types of orders.

\vspace{-0.3cm}
\subsubsection{Chirality checks}
\vskip -0.2cm

An additional analysis is performed in Fig.~\ref{FigSM:VBS_Chirality}(b) for $\Delta\!=\!1.2$ and $J_2\!=\!0.1$, the same $20\!\times\!6$ non-scan with the $S^z$ conserved as before, to detect the signs of the scalar chiral order in the SL state. We have biased the system towards the chiral-broken state by introducing a chiral term ${\cal H}\!=\!{\bf S}_i\cdot({\bf S}_j\!\times\!{\bf S}_k)$ in the triangles of the rightmost column of the cylinder and, separately, in one triangle at center.

Fig.~\ref{FigSM:VBS_Chirality}(b) shows the exponential decay from the pinned chirality at the right boundary (black symbols and their fit by the blue line) with the correlation length $\sim\!1.3a$. The correlations also become isotropic toward the center of the cylinder. Fig.~\ref{FigSM:VBS_Chirality}(b) also shows a faster exponential decay away from the chiral pinning at the center (red symbols) with a somewhat non-trivial anisotropic pattern in the correlations. The slow removal of the chiral pinning from the center of the cluster, until reaching a chiral term of $\mathcal{O}(10^{-9})$ to ensure complex wavefunctions, yields the isotropic correlations and the chirality values between $10^{-6}$ and $10^{-5}$ at the center of the cluster, suggesting no chiral SL. 

\vspace{-0.3cm}
\subsubsection{Wider $20\times 9$ non-scan YC cylinder}
\vskip -0.2cm

\begin{figure}[h!]
\includegraphics[width=0.75\linewidth]{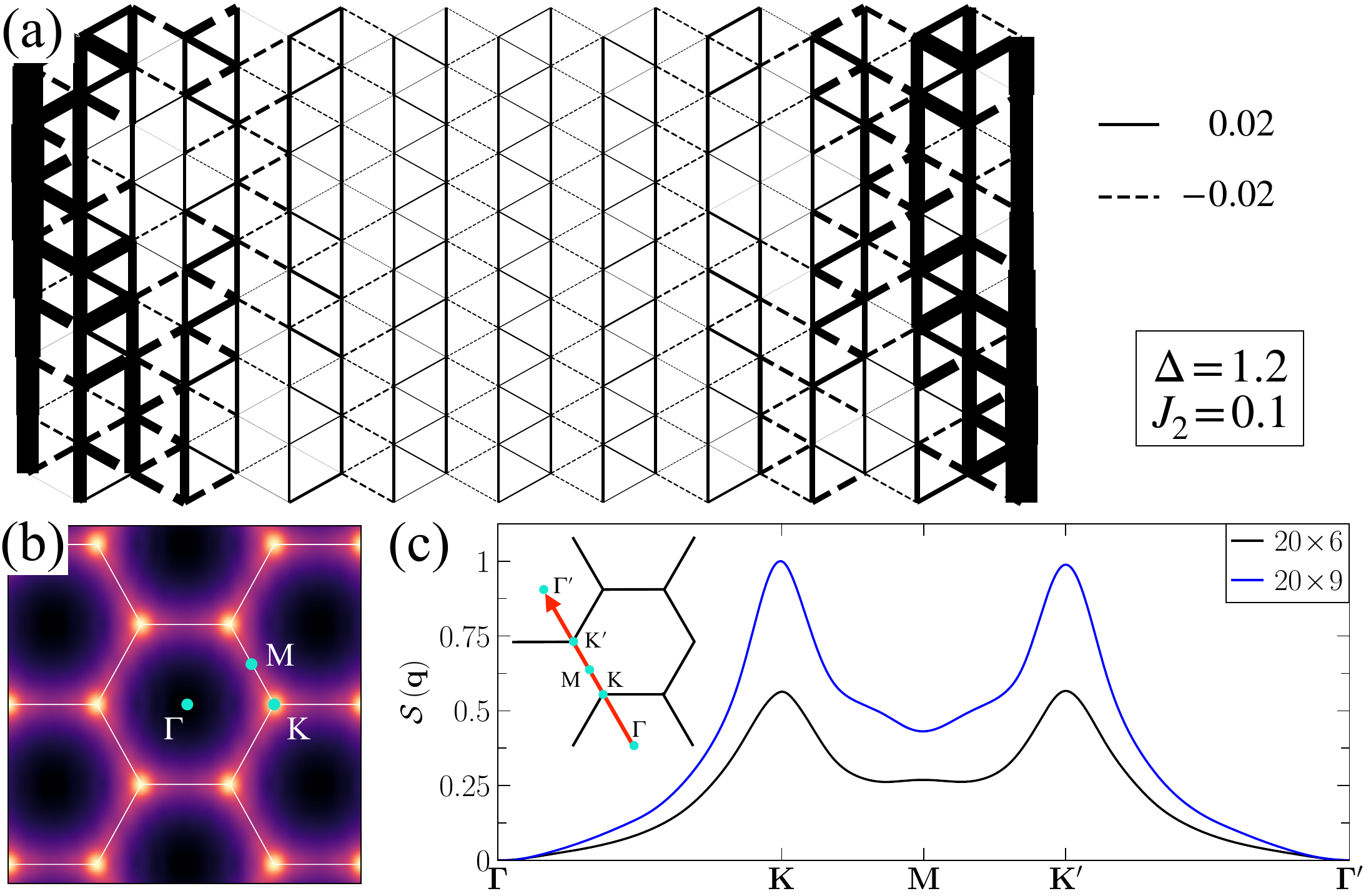}
\caption{All for $\Delta\!=\!1.2$ and $J_2\!=\!0.1$ in the $20\times 9$ non-scan YC cylinder. (a) The NN correlators with subtracted average. (b) Same as in Fig.~\ref{Fig:SLPLots}(b),  $\mathcal{S}({\bf q})$ in the SL phase. (c) Same as in Fig.~\ref{Fig:SLPLots}(d) with $\mathcal{S}({\bf q})$ from the $20\!\times\!6$ non-scan for comparison.}
\label{FigSM:YC9}
\end{figure}

Fig.~\ref{FigSM:YC9} shows our results from the wider $20\!\times\!9$ cylinder for $\Delta\!=\!1.2$ and $J_2\!=\!0.1$ in the SL phase, without   pinning fields, $S^z$ conserved, and keeping up to $m\!=\!8000$ states. The weakly anisotropic NN correlations in Fig.~\ref{FigSM:YC9}(a) are a signature of   degenerate ground states present in the $L_y$-odd cylinders that are also observed in the isotropic Heisenberg case~\cite{Zhu2015J1J2}.  Fig.~\ref{FigSM:YC9}(b) shows  $\mathcal{S}({\bf q})$, which exhibits the same features of the SL state as in the $20\!\times\!6$ cluster; see Fig.~\ref{Fig:SLPLots}(b).  Fig.~\ref{FigSM:YC9}(c) shows $\mathcal{S}({\bf q})$ along the ${\bf q}$-path shown in the inset, also compared to the one from the $20\!\times\!6$ cluster for the same parameters; both  are normalized by the maximum at ${\bf q}\!=\!K$ in the $20\!\times9$ cluster.

\vspace{-0.3cm}
\subsection{Fig.~\ref{Fig:DMRGPHD} data}
\vskip -0.2cm

The Tables below present a compilation of the DMRG data for the phase boundaries that are  used in Figs.~\ref{Fig:DMRGPHD} and \ref{FigEM:NonScanPHD}(a).  Table~\ref{tabSM:easy-axis} provides the transition points from both the scans and non-scans in the easy-axis version of model~\eqref{eq:spin_hamiltonian} obtained in this work. Table~\ref{tabSM:easy-plane} presents the transition points in the easy-plane version of model~\eqref{eq:spin_hamiltonian} from the scans, obtained in Ref.~\cite{Zhu2017YMGO}.

\begin{table}[h!]
\setlength{\arrayrulewidth}{0.2mm}
\setlength\extrarowheight{2.5pt}
\setlength{\tabcolsep}{3pt}
\begin{tabular}{ c || c | c | c | c | c || c | c | c }
\hline \hline 
\ Transition \ & \multicolumn{5}{c||}{\ Scan data $(\Delta, J_2)$\ } & \multicolumn{3}{c}{\ Non-scan data $(\Delta, J_2)$\ } \\ 
\hline \hline

\ Y --- SL \  & $(1,\ 0.055)$  & $(1.1,\ 0.057)$ &  $(1.3,\ 0.062)$  &  $(1.6,\ 0.073)$  &  $(1.9,\ 0.082)$  & $(1,\ 0.068)$  & $(1.1,\ 0.070)$ &  $(1.3,\ 0.076)$  \\ \hline

\ SL --- stripe-z \  & $(1,\ 0.157)$  & $(1.1,\ 0.153)$ &  $(1.3,\ 0.140)$  &  $(1.6,\ 0.117)$  &  $(1.9,\ 0.099)$ & $(1,\ 0.140)$  & $(1.1,\ 0.130)$ &  $(1.3,\ 0.115)$  \\ \hline

\ \ Y --- stripe-z \ & $(2.5,\ 0.088)$  & $(3.3,\ 0.072)$  & $(5,\ 0.049)$  &  $(10,\ 0.024)$  &  $(\infty,\ 0)$ & $(2.5,\ 0.086)$ & $(5,\ 0.048)$  &  \\ \hline

\end{tabular}
\caption{DMRG scan and non-scan transition points for the easy-axis model~\eqref{eq:spin_hamiltonian}.}
  \label{tabSM:easy-axis}
\end{table}
\vskip -0.3cm

\begin{table}[h!]
\setlength{\arrayrulewidth}{0.2mm}
\setlength\extrarowheight{2.5pt}
\setlength{\tabcolsep}{3pt}
\begin{tabular}{ c || c| c | c | c | c | c }
\hline \hline 
\ Transition \ & \multicolumn{5}{c}{\ Scan data $(\Delta, J_2)$\ } \\ 
\hline \hline

\ $120^\circ$ --- SL \ & $(1,\ 0.055)$ & $(0.8,\ 0.065)$  & $(0.7,\ 0.071)$  &  $(0.6,\ 0.078)$  &  $(0.5,\ 0.085)$   \\ \hline

\ SL --- stripe \ & $(1,\ 0.157)$  & $(0.8,\ 0.150)$ &  $(0.7,\ 0.144)$  &  $(0.6,\ 0.137)$  &  $(0.5,\ 0.130)$  \\ \hline

\ \ $120^\circ$ --- stripe \ & $(0.3,\ 0.103)$  & $(0,\ 0.007)$  \\ \hline

\end{tabular}
\caption{DMRG scan transition points for the easy-plane model~\eqref{eq:spin_hamiltonian} from Ref.~\cite{Zhu2017YMGO}.}
\label{tabSM:easy-plane}
\end{table}
\vskip -0.2cm

\vspace{-0.3cm}
\subsection{More details on the transitions}

\vspace{-0.3cm}
\subsubsection{More  $\mathcal{S}({\bf q})$}
\vskip -0.2cm

\begin{figure}[h!]
\includegraphics[width=0.55\linewidth]{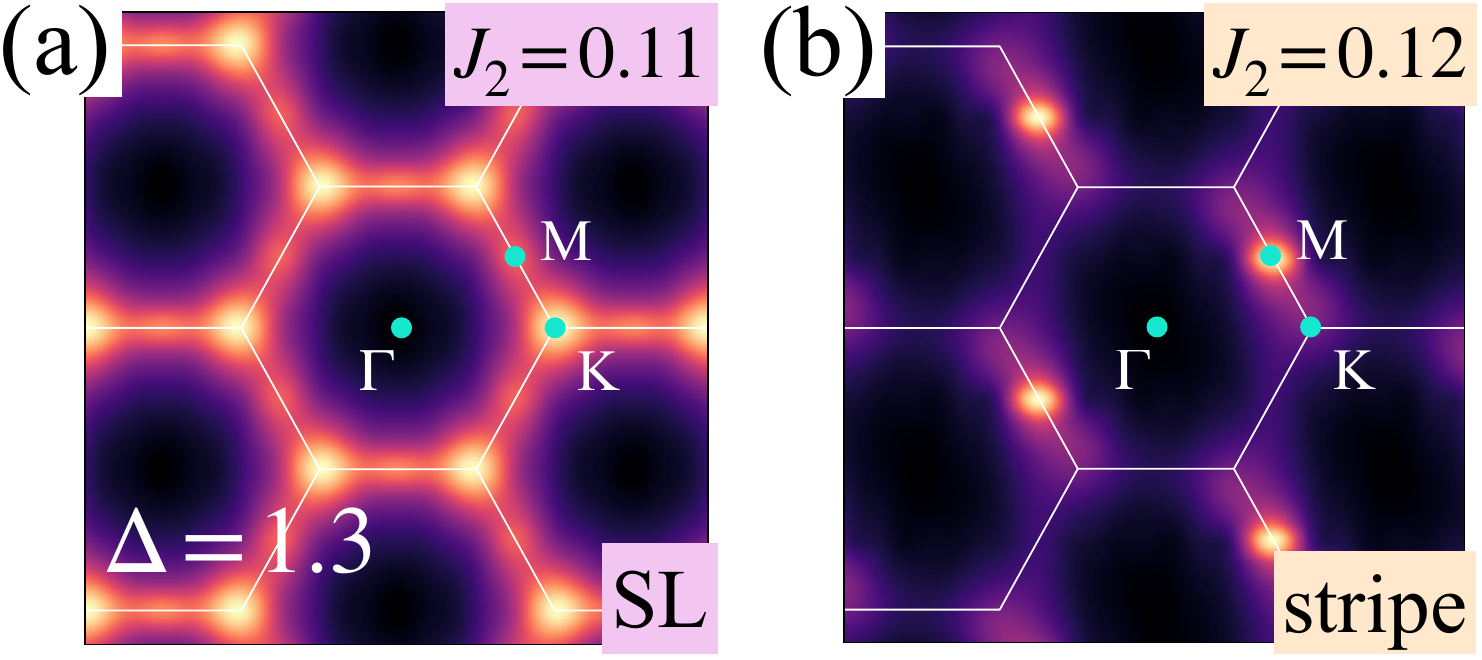}
\vskip -0.3cm
\caption{$\mathcal{S}({\bf q})$ intensity plots for $\Delta\!=\!1.3$, same as Figs.~\ref{Fig:SLPLots}(b) and \ref{Fig:SLPLots}(c), for (a)  $J_2\!=\!0.11$ and  (b) $J_2\!=\!0.12$.}
\label{FigSM:Sq_M}
\vskip -0.3cm
\end{figure}

In Fig.~\ref{FigSM:Sq_M}, $\mathcal{S}({\bf q})$ intensity plots for $\Delta\!=\!1.3$ and (a) $J_2\!=\!0.11$ (SL phase) and (b) $J_2\!=\!0.12$ (stripe-z phase) are shown. They are obtained from the non-scan $20\times 6$ YC cylinders without pinning field, $S^z$ not conserved, and $m$ up to 2500. The $\mathcal{S}({\bf q})$  profiles in Fig.~\ref{FigEM:NonScanPHD}(b)  for these values of $J_2$ used the same data.

\vspace{-0.3cm}
\subsubsection{Direct Y-to-stripe-z transition}
\vskip -0.2cm

\begin{figure}[h!]
\includegraphics[width=.85\linewidth]{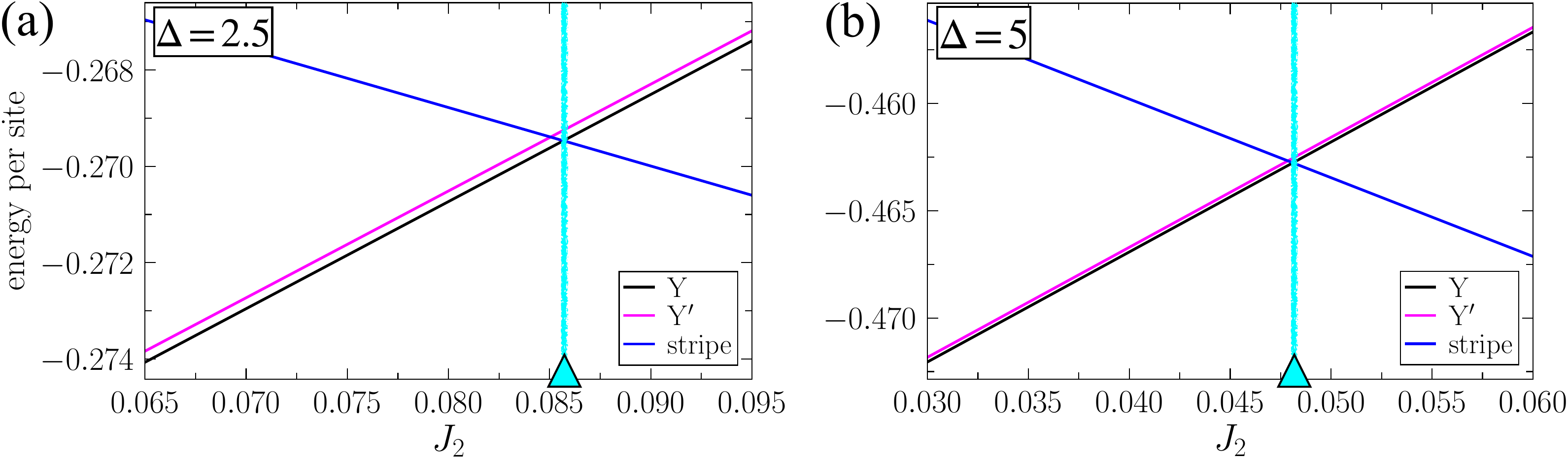}
\vskip -0.3cm
\caption{Energies of the three competing phases vs $J_2$. Lines are extrapolated energies, $\langle \psi_i | \hat{\mathcal{H}}(J_2) |\psi_i\rangle$, where $|\psi_i\rangle $ are the Y and Y$'$ states at $J_2\!=\!0.065$ ($0.03$), and stripe-z at $J_2\!=\!0.095$ ($0.06$) for $\Delta\!=\!2.5$ ($5.0$).}
\label{FigSM:EnergyCrossing}
\vskip -0.3cm
\end{figure}

For the first-order Y-to-stripe-z transitions for $\Delta\!\agt\!2.0$, the transition points can be alternatively obtained from the crossings of the DMRG energies of the competing states using extrapolations based on the spin-spin correlations extracted from the center of the non-scan clusters for each of the states. Triangles in Fig.~\ref{FigSM:EnergyCrossing} mark such crossings for $\Delta\!=\!2.5$ and $5.0$ along the $J_2$ axis. This figure shows the extrapolated energies, $\langle \psi_i | \hat{\mathcal{H}}(J_2) |\psi_i\rangle$,  for the three competing states, stripe-z and the near-degenerate Y and Y$^\prime$ phases, with the states $|\psi_i\rangle $ for the Y and Y$'$ states evaluated at $J_2\!=\!0.065$ ($0.03$) and for the stripe-z at $J_2\!=\!0.095$ ($0.06$) for $\Delta\!=\!2.5$ ($5.0$), respectively. The Y state is always lower than the Y$'$ state. The transition points obtained this way coincides with the ones from the inflection points of the order parameters in the DMRG scans, such as the one in Fig.~\ref{Fig:YtoStripe} of the main text, within the error bars.

\vspace{-0.3cm}
\subsubsection{Narrower $J_2$-scan for the Y-to-stripe-z boundary}
\vskip -0.2cm

\begin{figure}[h!]
\includegraphics[width=.7\linewidth]{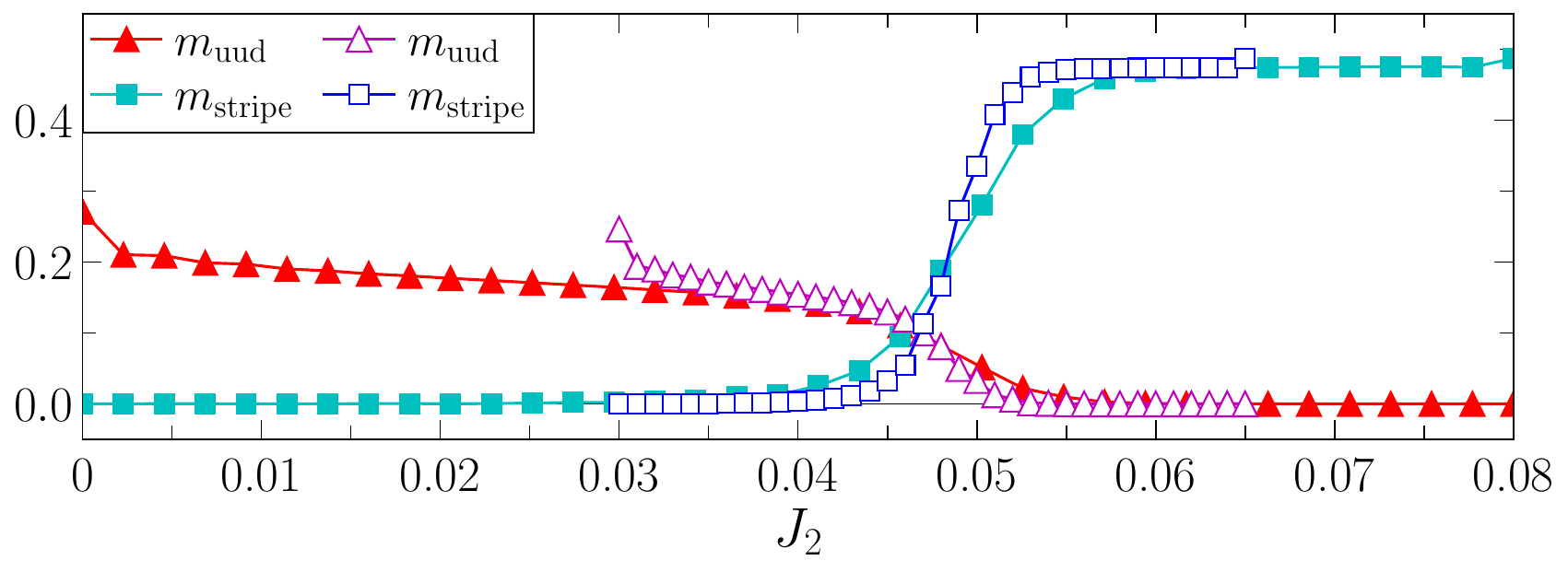}
\vskip -0.3cm
\caption{The $m_\mathrm{uud}$ and $m_\mathrm{stripe}$ order parameters vs $J_2$ for $\Delta\!=\!5$ and two different ranges of the scan, c.f., Fig.~\ref{Fig:YtoStripe}(b).}
\label{FigSM:YtoStripe}
\vskip -0.3cm
\end{figure}

In Fig.~\ref{FigSM:YtoStripe}, the results for $m_\mathrm{uud}$ and $m_\mathrm{stripe}$ order parameters from the $J_2$-scan for $\Delta\!=\!5$ in Fig.~\ref{Fig:YtoStripe}(b) are superimposed with the same results for the narrower $J_2$-scan in order to demonstrate that the Y-to-stripe-z transition width narrows down together with the scan, supporting its first-order character. The transition inferred  from the inflection points of the order parameters closely coincides with the energy-crossing in the non-scans, discussed above.

\vspace{-0.3cm}
\subsubsection{$J_2$-scan in the Ising limit}
\vskip -0.2cm

\begin{figure}[h!]
\includegraphics[width=\linewidth]{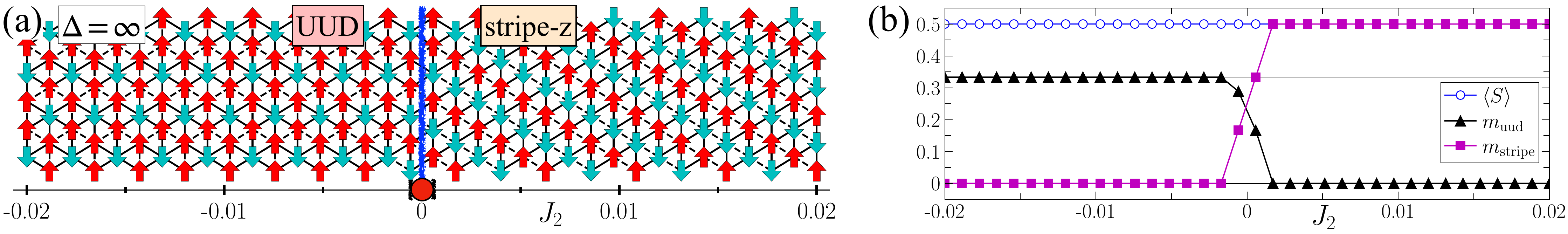}
\vskip -0.3cm
\caption{(a) The $J_2$-scan in the Ising limit, $\Delta\!=\!\infty$. (b) Order parameters vs $J_2$.}
\label{FigSM:IsingScan}
\vskip -0.3cm
\end{figure}

 Figure~\ref{FigSM:IsingScan} shows the $J_2$-scan in the Ising limit,  going through the degenerate $J_2\!=\!0$ point and showing a direct transition from the {\it classical} UUD to the stripe-z phase upon changing the sign of $J_2$. Colors of the arrows in Fig.~\ref{FigSM:IsingScan}(a) are for the $\pm S^z\!=\!1/2$. The horizontal line in Fig.~\ref{FigSM:IsingScan}(b) shows the $1/3$ magnetization of the UUD state. 

\vspace{-0.3cm}
\subsection{$1/L_y$-scaling of the supersolid order parameters}
\vskip -0.2cm

\begin{figure}[h!]
\includegraphics[width=0.8\linewidth]{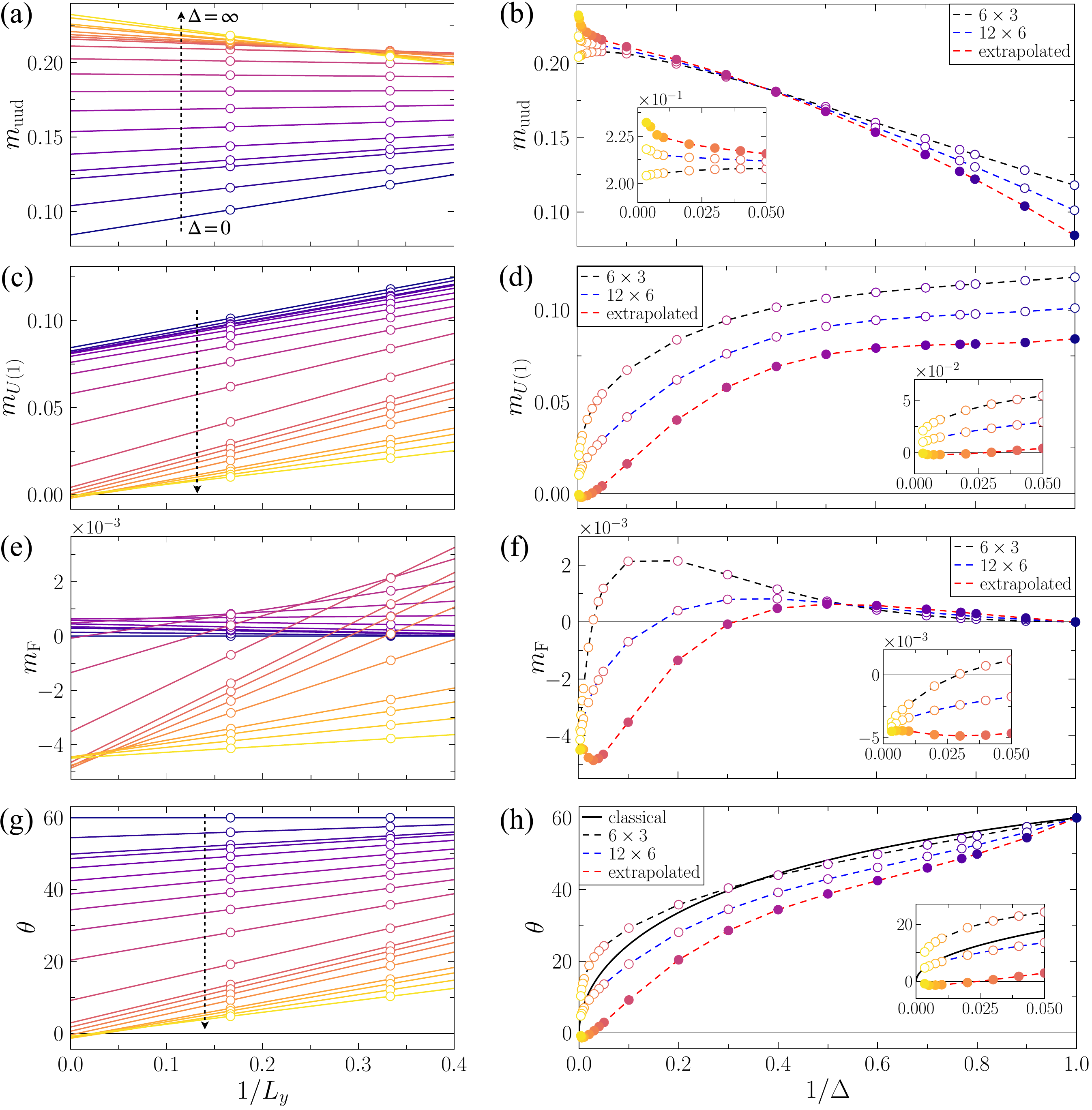}
\vskip -0.3cm
\caption{The $1/L_y$-scaling of the different order parameters and the tilt angle for $J_2\!=\!0$ and different values of $\Delta$.}
\label{FigSM:OrderParametersScalingJ20}
\vskip -0.4cm
\end{figure}

To investigate the supersolid order parameters of the Y-phase, we have used the $1/L$-scaling in fixed aspect ratio DMRG clusters, described in the main text and EM for the ordered moment $\langle S\rangle$, with the results for  $J_2\!=\!0$ vs $1/\Delta$  presented in Fig.~\ref{Fig:OrderParameters} in the main text. In Fig.~\ref{FigSM:OrderParametersScalingJ20}, we show the details of these results.

Figs.~\ref{FigSM:OrderParametersScalingJ20}(a), (c), (e), and (g) show the linear extrapolation using  $6\!\times \!3$ and $12\!\times \!6$ clusters with the edges pinned by the classical Y phase for the $m_\mathrm{uud}$, $m_{U(1)}$,  $m_\mathrm{F}$ order parameters, and the spin tilt angle $\theta$ in the Y phase, respectively. Figs.~\ref{FigSM:OrderParametersScalingJ20}(b), (d),  (f), and (h) show the values of these orders and the angle in the centers of the clusters together with their extrapolated values vs $1/\Delta$ from the Ising to the Heisenberg  limit. One can see that the extrapolation is becoming more problematic near the Ising limit, likely because of the smaller values of  $m_{U(1)}$. However, the ``solid'' order parameter $m^\infty_{\rm uud}$ is clearly finite at $\Delta\!\rightarrow\!\infty$, while $m^\infty_{U(1)}$ is more likely not. The scale for $m_\mathrm{F}$ component is at least an order of magnitude smaller for most $1/\Delta$, with non-zero values that are showing the logical limit for the reliability of the linear extrapolation procedure. 

\begin{figure}[h!]
\includegraphics[width=0.8\linewidth]{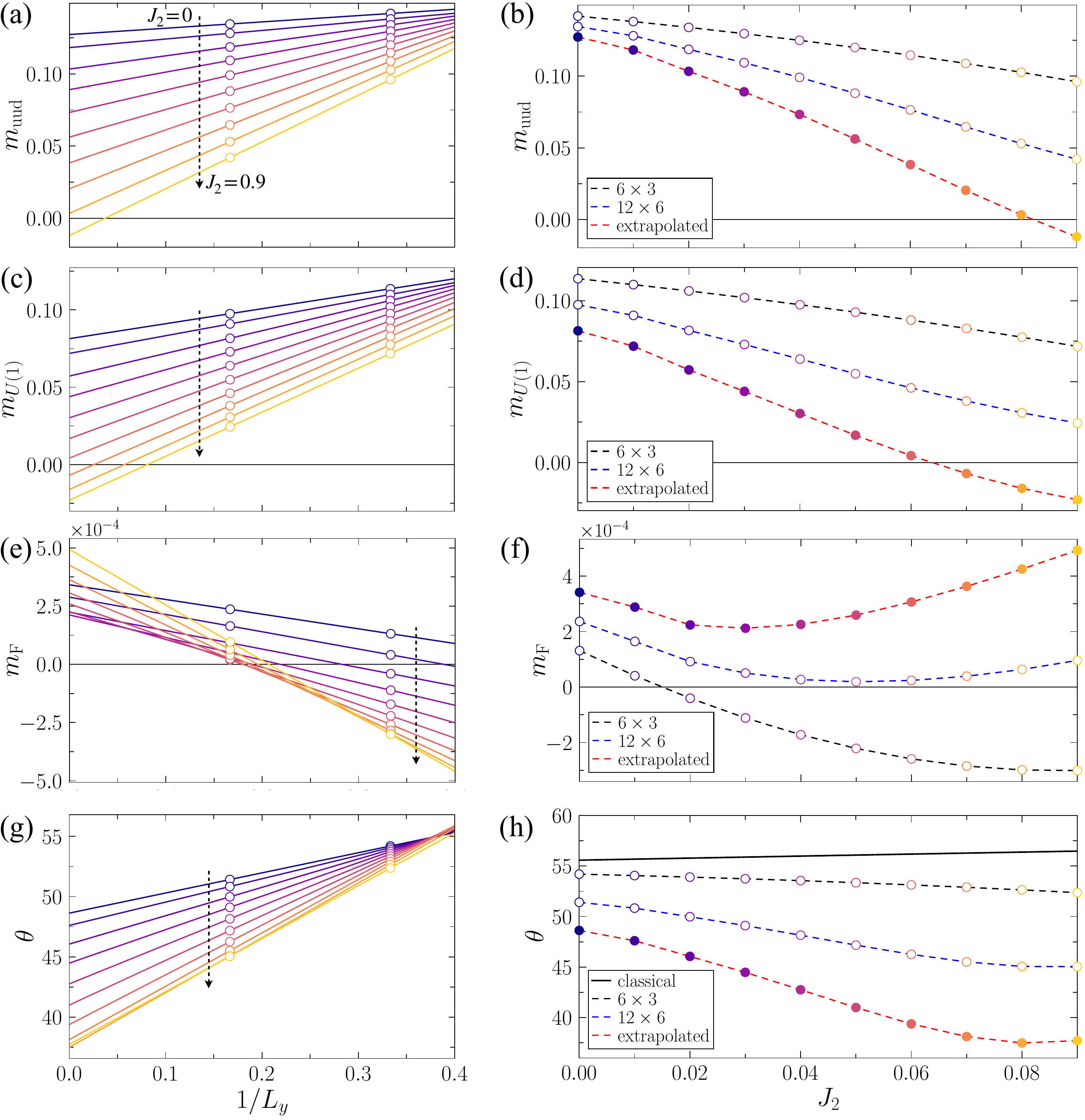}
\vskip -0.3cm
\caption{Same as Fig.~\ref{FigSM:OrderParametersScalingJ20} for $\Delta\!=\!1.3$ and different values of $J_2$.}
\label{FigSM:OrderParametersScalingD1p3}
\end{figure}

We have also performed similar analysis of the order parameters  for several fixed values of $\Delta$ and a set of $J_2$. In  Fig.~\ref{FigSM:OrderParametersScalingD1p3}, we show the same $1/L$-scaling as in Fig.~\ref{FigSM:OrderParametersScalingJ20} for $\Delta\!=\!1.3$ vs $J_2$. The $1/L$-scaling in Fig.~\ref{FigSM:OrderParametersScalingD1p3} may suggest that the extrapolated $m^\infty_{U(1)}$ order vanishes somewhat before $m^\infty_{\rm uud}$, one undershooting and the other overshooting the Y-to-SL phase boundary. We have tracked this behavior for several values of $\Delta$. While the existence of a thin layer of the pure up-up-down state with $m_{U(1)}\!=\!0$ in the vicinity of the border to the SL and stripe-z phases cannot be ruled out,  most likely scenario is the  non-linear effects in the finite-size extrapolations for the already small values of  $m_{U(1)}$. One should also point out that while the $m^\infty_{U(1)}$ order parameter extrapolates to zero and negative values close to the phase boundary, the tilt angle $\theta$ does not.

\section{Quasiclassical analysis}

\vspace{-0.2cm}
\subsection{Classical phase transition}
\vskip -0.2cm

Using the classical energies of the Y$^\prime$ and stripe-z phases in EM,  the transition between them is given by
\begin{equation}
    \bar{\Delta}=\frac{\sqrt{J_{2} (1+J_{2}) (3-8 J_{2} (1-2 J_{2}))}+2 J_{2} (1+J_{2})}{ (1-2J_{2}) (1+J_{2})},
\label{eqSM:YptoStz}
\end{equation}
which yields the square-root dependence for small $J_2\!>\!0$ near the Ising limit, $\bar{\Delta} \!\approx \!\sqrt{3J_{2}}$.

Similarly, the classical transition between the Y and UUD phases reads: $\bar{\Delta}\!=\!-2J_2/(1-2J_2)$, readily yielding the linear dependence  for small $J_2\!<\!0$  near the Ising limit, $\bar{\Delta} \!\approx \!-2J_{2}$.

For the easy-plane region, the classical transition line between the $120^\circ$ and stripe states is $J_2\!=\!1/8$ for all $\Delta\!<\!1$~\cite{Zhu2017YMGO}.

\vspace{-0.3cm}
\subsection{Spin-wave theory}

For the $1/S$ spin-wave expansion, the laboratory reference frame $\{x, y, z\}$ needs to be rotated to the local reference frame $\{\tilde{x}, \tilde{y},\tilde{z}\}$ at each site, so that $\tilde{z}$ aligns with the direction of the spins given by the state that minimizes the classical energy.

For the Y state, with the spins in the $x$-$z$ plane, the spin components $\widetilde{\bf S}$ in the local reference frame are related to those in the laboratory frame according to $ S_{A(B,C)}^{y}=\widetilde{S}_{A(B,C)}^y$,  $\big(S_A^x, S_A^z\big)\!=\!-\big(\widetilde{ S}_A^x, \widetilde{S}_A^z\big)$, and 
\begin{align}
\big\{S_{B(C)}^{x},S_{B(C)}^{z}\big\}=\big\{\widetilde{S}_{B(C)}^x\cos \theta \pm \widetilde{S}_{B(C)}^z\sin \theta, \widetilde{S}_{B(C)}^z\cos \theta \mp \widetilde{S}_{B(C)}^x\sin \theta\big\},
\label{eqSM:SlocY}
\end{align}
with the angle $\theta$ given in EM of the main text by Eq.~\eqref{eqEM:thetaclY}. 

Similarly, for the Y$^\prime$ state, the transformation is given by $ S_{A(B,C)}^{y}=\widetilde{S}_{A(B,C)}^y$,  $\big(S_A^x, S_A^z\big)\!=\!\big(\! -\!\widetilde{ S}_A^z, \widetilde{S}_A^x\big)$, and
\begin{align}
\big\{S_{B(C)}^{x},S_{B(C)}^{z}\big\}=\big\{\pm \widetilde{S}_{B(C)}^x\cos \theta + \widetilde{S}_{B(C)}^z\sin \theta,\pm \widetilde{S}_{B(C)}^z\cos \theta - \widetilde{S}^{x}_{B(C)}\sin \theta\big\},
\label{eqSM:SlocYp}
\end{align}
with the angle $\theta$ given in EM by Eq.~\eqref{eqEM:thetaclYp}. 

\subsubsection{Linear spin-wave theory}

The linear spin-wave theory (LSWT) order of the $1/S$-expansion about the classical ground state is obtained via the standard Holstein-Primakoff (HP) bosonization of the spin operators in the local reference frame: $\widetilde{S}_{\nu, \ell}^{z}=S-n_{\nu, \ell}$, with $n_{\nu, \ell}\!=\!a^\dag_{\nu, \ell}a_{\nu, \ell}$, and, to the lowest order, $\widetilde{S}_{\nu, \ell}^{+}\approx \sqrt{2S} a_{\nu, \ell}$, where we use the notations of the magnetic unit cell $\ell$ and sublattice index $\nu$. 

Specifically, for the three-sublattice orders in the TL, one needs  three species of bosons, $a_{\nu, \ell}=\{ a_\ell, b_\ell, c_\ell\}$, in the HP transformation. With the Fourier transformation
\begin{equation}
a_{\nu, \ell}= \frac{1}{\sqrt{N_c}}\sum_{\bf q}a_{\nu , {\bf q}}e^{-i{\bf q} {\bf r}_{\nu, \ell}},
\end{equation}
where $N_c=N/3$ is the number of unit cells, the LSWT Hamiltonian in the TL for the three bosonic flavors can be written in the matrix form
\begin{equation}
\hat{\mathcal{H}}^{(2)} = \frac{3S}{2} \sum_{\bf q} \hat{{\bf x}}_{\bf q}^\dagger \hat{{\bf H}}_{\bf q}\hat{{\bf x}}_{\bf q}^{\phantom\dag},\quad
\hat{{\bf H}}_{\bf q} = \begin{pmatrix}
\hat{{\bf A}}_{\bf q} & \hat{{\bf B}}_{\bf q} \\
\hat{{\bf B}}_{\bf q}^\dagger & \hat{\bf A}^*_{-{\bf q}} \end{pmatrix},  
\label{eqSM:H2kYYp}
\ \mbox{with}\quad 
\hat{{\bf A}}_{\bf q}=\begin{pmatrix}
A_{\bf q} & D_{\bf q} & E_{\bf q}^* \\
D_{\bf q}^* & B_{\bf q} & F_{\bf q}\\
E_{\bf q} & F_{\bf q}^* & C_{\bf q}\\
\end{pmatrix}, \ \mbox{and}\quad 
\hat{{\bf B}}_{\bf q}=\begin{pmatrix}
G_{\bf q} & J_{\bf q} & K_{\bf q}^* \\
J_{\bf q}^* & H_{\bf q} & L_{\bf q}\\
K_{\bf q} & L_{\bf q}^* & I_{\bf q}\\
\end{pmatrix},  
\end{equation}
and $\hat{{\bf x}}_{\bf q}^\dagger \!=\! \big( a_{\bf q}^\dagger, b_{\bf q}^\dagger, c_{\bf q}^\dagger, a_{-{\bf q}}^{\phantom\dag}, b_{-{\bf q}}^{\phantom\dag}, c_{-{\bf q}}^{\phantom\dag} \big)$ are the bosonic vector operators. 

For the Y phase, the elements of the $3\times 3$ matrices $ \hat{{\bf A}}_{\bf q}$ and $ \hat{{\bf B}}_{\bf q}$ are 
\begin{align}
&A_{\bf q} \!=\! 2 \Delta \cos \theta\!+\! 2J_2  (\gamma^{(2)}_{\bf q} \!-\! \Delta) , \ \ \ B_{\bf q} \!= C_{\bf q} \!= \sin^2\!\theta\!+\!\Delta(1\!-\!\cos\theta)\cos\theta 
\!+\! J_2\big(2\gamma_{\bf q}^{(2)} \!-\!2\Delta \!+\!(2\!+\!\gamma_{\bf q}^{(2)})(\Delta-1)  \sin^2\theta\big),
\nonumber\\ 
&D_{\bf q} =E_{\bf q} = \gamma_{\bf q}^{\phantomsection} \sin^2(\theta/2) ,\quad\quad F_{\bf q} =\frac{\gamma_{\bf q}^{\phantomsection}}{2}\big( 1+\cos^2\theta -\Delta \sin^2\theta \big) , 
\label{eqSM:AkMatrixY}\\
&G_{\bf q} =0,\quad H_{\bf q} = I_{\bf q} = J_2  \gamma_{\bf q}^{(2)} (\Delta -1) \sin ^2\theta,\quad J_{\bf q} = K_{\bf q} = -\frac{\gamma_{\bf q}^{\phantomsection}}{2}  (1+\cos \theta),\quad L_{\bf q} \!=\!  -\frac{\gamma_{\bf q}^{\phantomsection}}{2}   (\Delta+1)\sin ^2\theta,
\label{eqSM:BkMatrixY}
\end{align}
where $\gamma_{\bf q}\!=\!\frac{1}{3}\sum_{\alpha}e^{i{\bf q} {\bm \delta}_\alpha}$ and $\gamma_{\bf q}^{(2)}\!=\!\frac{1}{3}\sum_{\alpha}\cos ({\bf q} \tilde{{\bm \delta}}_\alpha)$, with the first- and second-neighbor translation vectors, ${\bm \delta}_\alpha$ and $\tilde{{\bm \delta}}_\alpha$, respectively. 

Similarly, for the Y$^\prime$, the elements of $ \hat{{\bf A}}_{\bf q}$ and $ \hat{{\bf B}}_{\bf q}$ read 
\begin{align}
&A_{\bf q}\! =\! 2\sin \theta\!+ \!J_2\big(  \gamma^{(2)}_{\bf q} (1\!+\! \Delta)\!-\!2\big), \quad B_{\bf q} \!= C_{\bf q}\! = \sin\theta\!-\!\sin^2\theta\!+\!\Delta \cos^2\theta\!
+\! J_2\big(2\gamma_{\bf q}^{(2)} \!-\!2\Delta \!+\!(2\!+\!\gamma_{\bf q}^{(2)})(\Delta-1)  \sin^2\theta\big)\nonumber, \\
&D_{\bf q} =E_{\bf q} = \frac{\gamma_{\bf q}^{\phantomsection}}{2} (1-\Delta\sin\theta) ,\quad\quad F_{\bf q} =\frac{\gamma_{\bf q}^{\phantomsection}}{2}( 1+\Delta) \sin^2\theta, 
\label{eqSM:AkMatrixYp} \\
&G_{\bf q} \!=\!J_2\gamma_{\bf q}^{(2)}(\Delta\!-\!1),\ \  H_{\bf q} \!=\! I_{\bf q} \!= \!J_2  \gamma_{\bf q}^{(2)} (\Delta \!-\!1) \sin ^2\theta,\ \ J_{\bf q} \!=\! K_{\bf q}\! = \!-\frac{\gamma_{\bf q}^{\phantomsection}}{2}  (1\!+\!\Delta \sin \theta),\ \ L_{\bf q} \!=\! \frac{\gamma_{\bf q}^{\phantomsection}}{2}   \big((1\!+\!\Delta)\sin ^2\theta\!-\!2\big).
\label{eqSM:BkMatrixYp}
\end{align}
The eigenvalue problem of Eq.~\eqref{eqSM:H2kYYp}, is solved by diagonalizing $(\hat{\bf g}\hat{ \bf H}_{\bf q})$ to obtain the magnon eigenenergies $\varepsilon_{\nu,{\bf q}}$, with the para-unitary diagonal matrix $\hat{\bf g}\!=\![1,1,1,-1,-1,-1]$. 

For the stripe-z phase, the LSWT Hamiltonian can be written  in terms of the single bosonic species 
\begin{equation}
\mathcal{H}^{(2)}=2S\sum_{\bf q} \left( \bar{A}_{\bf q}a^{\dagger}_{\bf q}a^{\phantomsection}_{\bf q}-\frac{{\bar B}_{\bf q}}{2}\big( a^{\dagger}_{\bf q}a^{\dagger}_{-{\bf q}} + a_{-{\bf q}}a_{{\bf q}} \big)\right),
\label{eqSM:H2kSt}
\end{equation}
where ${\bar A}_{\bf q} \!=\! (1\!+\!J_2)\Delta+\cos({\bf q} {\bm \delta}_{1})+J_2\cos({\bf q} \tilde{\bm \delta}_{2})$, and  ${\bar B}_{\bf q} \!=\! \cos({\bf q} {\bm \delta}_{2})\!+\!\cos({\bf q} {\bm \delta}_{3})+J_2\big(\cos({\bf q} \tilde{\bm \delta}_{1})+\cos({\bf q} \tilde{\bm \delta}_{3})$\big). The LSWT Hamiltonian~\eqref{eqSM:H2kSt} is diagonalized by a textbook Bogolyubov transformation, $a_{\bf q}=u_{\bf q}d_{\bf q}^{\phantomsection} + v_{\bf q}d_{-{\bf q}}^\dagger$, with $u_{\bf q}^2+v_{\bf q}^2=\bar{A}_{\bf q}/\varepsilon_{\bf q}$, $2u_{\bf q}v_{\bf q}=\bar{B}_{\bf q}/\varepsilon_{\bf q}$, and the magnon energy $\varepsilon_{\bf q}\!=\!\sqrt{{\bar A}_{\bf q}^2-{\bar B}_{\bf q}^2}$. 

\subsubsection{Minimally augmented spin-wave theory}

Using the LSWT magnon energies, the leading $1/S$ quantum correction to the classical ground state energy is 
\begin{equation}
\delta E = \frac{1}{2}\sum_{\bf q}\Big(\sum_{\nu}\varepsilon_{\nu {\bf q}}-\mathrm{tr}\big( \hat{\bf A}_{\bf q} \big) \Big),
\label{eqSM:dEcl}
\end{equation} 
so that the $\mathcal{O}(S)$ energy of a given state is $E=E_{cl}+\delta E$.

\begin{figure}[t]
\includegraphics[width=.9\linewidth]{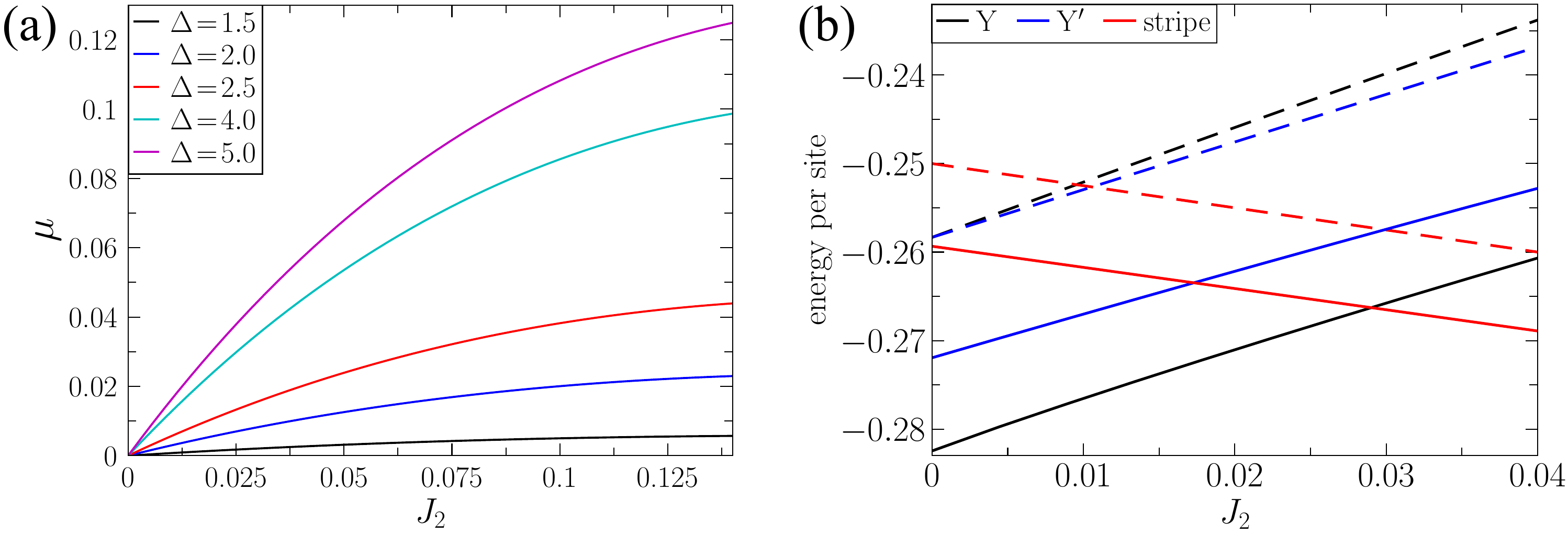}
\vskip -0.3cm
\caption{(a) The minimal value of $\mu(\Delta, J_2)$~(\ref{eqSM:mu}) vs $J_2$ for different values of $\Delta$. (b) The classical and quantum per site energies~\eqref{eqSM:dEcl}, dashed and solid lines, respectively,  for the Y, Y$^\prime$, and stripe-z states vs $J_2$ for $\Delta\!=\!5$.}
\label{FigS:MAGSWT}
\vskip -0.4cm
\end{figure}

However,  for $J_2\!>\!0$,  the Y state is not  the minimum  of the classical energy, being overcome by the Y$'$ state, as is shown by the dashed lines in Fig.~\ref{FigS:MAGSWT}(b),  and the LSWT diagonalization of the Hamiltonian in~\eqref{eqSM:H2kYYp} for the Y state yields the unphysical (complex)  magnon energies $\varepsilon_{\mu, {\bf q}}$ near the $\Gamma$-point, such that the $1/S$  correction in~\eqref{eqSM:dEcl} becomes ill-defined. 

The minimally augmented spin-wave theory (MAGSWT) helps to resolve this problem by stabilizing the spectrum using  a local field in the direction of the ordered moments ${\bf n}_i$ of the classical spin configuration in the form of $ \delta \hat{\mathcal{H}}\!=\!\mu \sum_i(S-{\bf S}_i\cdot {\bf n}_i)$~\cite{Wenzel2012MAGSWT, Colleta2013MAGSWT, Jiang2023J1J3Honeycomb}. This is equivalent to introducing a positive shift in the chemical potential of the bosonic HP operators, while leaving the classical energy of the state unchanged. The value of the chemical potential is chosen to ensure that the spectrum is positively defined everywhere in the Brillouin zone. With the corrected spectrum, the $1/S$  correction to the energy can be calculated in the standard form~\eqref{eqSM:dEcl}.

For the Y state, the LSWT Hamiltonian in Eq.~\eqref{eqSM:H2kYYp} can be analytically diagonalized at the $\Gamma$-point and the minimal value of $\mu(\Delta, J_2)$, shown in Fig.~\ref{FigS:MAGSWT}(a) vs $J_2$ for different values of $\Delta$, can be derived as
\begin{align}
\mu(J_2,\Delta)&\!=\!\frac{3S\cos^2\theta}{2\Delta^2}\Big(2\big(1+\Delta
+J_2(\Delta-1)\big)\big(2J_2(\Delta-1)^2-\Delta^2\big)-1-\Delta+\sqrt{\lambda}\Big),
\label{eqSM:mu}
\end{align}
where
\begin{align}
&\lambda\!=\! 256 J_2^6 (\Delta \!-\!1)^6 \!+\!128 J_2^5 (\Delta\!-\!1)^5 (5\!+\!3\Delta) \!+\! J_2^4 (\Delta\!-\!1)^4 \big( 656\!+\!16\Delta (46\!-\!3\Delta)\big)\!+\!J_2^3(\Delta\!-\!1)^3 \big[352\!+\!16\Delta\big(34\!-\!\Delta(7\!+\!13\Delta))\big)\!\big]  \nonumber\\
&+\!J_2^2 (\Delta\!-\!1)^2 \big[104\!+\!4\Delta\big(48\!-\!\Delta(14\!+\!\Delta(56\!+\!3\Delta))\big) \big]  \!+\!4J_2 (\Delta \!-\!1) (\Delta \!+\!1) \big( 4\!+\!\Delta(2\Delta^2\!-\!1)(3\Delta\!-\!4) \big) \!+\!(\Delta \!+\!1)^2 (2\Delta^2 \!+\! 1)^2.
\end{align}

The black solid line in  Fig.~\ref{FigS:MAGSWT}(b) shows the quantum $\mathcal{O}(S)$ energies  of the Y state vs $J_2$ for $\Delta\!=\!5$  obtained following the MAGSWT strategy. The other states have their magnon energies stable  throughout the extent of $J_2$ in Fig.~\ref{FigS:MAGSWT}(b)  and their quantum energy corrections are calculated using the standard $1/S$ approach of Eq.~\eqref{eqSM:dEcl}, without the MAGSWT intervention.

 This Figure underscores the main success of the MAGSWT, as the Y state has the lower quantum energy than the Y$'$ state. The crossings between the Y and stripe-z energies, such as  the one in Fig.~\ref{FigS:MAGSWT}(b), define the MAGSWT line for the phase transition, shown in the main text in Fig.~\ref{Fig:Quasiclassics} and in  EM in Fig.~\ref{FigEM:NonScanPHD}(a). This transition line terminates in Fig.~\ref{FigEM:NonScanPHD}(a) for  $1/\Delta\! \agt \!0.6$ due to the lack of the crossing of the quantum energy lines for the Y and stripe states in the ranges of $J_2$ where their excitations are well-defined.

\end{document}